\documentclass[english,draftclsnofoot,onecolumn,12pt]{IEEEtran}
\usepackage[T1]{fontenc}
\usepackage[latin9]{inputenc}
\usepackage{array}
\usepackage{float}
\usepackage{mathtools}
\usepackage{bm}
\usepackage{multirow}
\usepackage{amsmath}
\usepackage{amsthm}
\usepackage{amssymb}
\usepackage{graphicx}

\makeatletter

\newcommand{\noun}[1]{\textsc{#1}}
\providecommand{\tabularnewline}{\\}
\floatstyle{ruled}
\newfloat{algorithm}{tbp}{loa}
\providecommand{\algorithmname}{Algorithm}
\floatname{algorithm}{\protect\algorithmname}

\theoremstyle{plain}
\newtheorem{thm}{\protect\theoremname}
\theoremstyle{remark}
\newtheorem{rem}[thm]{\protect\remarkname}

\usepackage{graphicx, amsmath}
\usepackage{mathtools}
\usepackage{subfigure}
\usepackage{cite}
\usepackage{algorithmic}

\newcommand{\trans}{^{\mbox{\scriptsize T}}}
\newcommand{\herm}{^{\mbox{\scriptsize H}}}
\renewcommand{\fnum@figure}{Fig.~\thefigure}
\allowdisplaybreaks

\makeatother

\usepackage{babel}
\providecommand{\remarkname}{Remark}
\providecommand{\theoremname}{Theorem}

\begin{document}

\title{Energy Efficiency Fairness for Multi-Pair Wireless-Powered Relaying
Systems}

\author{Kien-Giang Nguyen,~\IEEEmembership{Student Member, IEEE}, Quang-Doanh
Vu,~\IEEEmembership{Member, IEEE}, Le-Nam Tran,~\IEEEmembership{Senior Member, IEEE}
and Markku Juntti,~\IEEEmembership{Senior Member, IEEE}\thanks{Manuscript received March 19, 2018; revised July 6, 2018; accepted
September 6, 2018. This work was supported in part by the Academy
of Finland under the projects \textquotedblleft Wireless Connectivity
for Internet of Everything\textendash Energy Efficient Transceiver
and System Design (WiConIE)\textquotedblright{} under Grant 297803,
\textquotedblleft Flexible Uplink-Downlink Resource Management for
Energy and Spectral Efficiency Enhancing in Future Wireless Net- works
(FURMESFuN)\textquotedblright{} under Grant 31089, and \textquotedblleft 6Genesis
Flagship\textquotedblright{} under Grant 318927. This publication
has emanated from research supported in part by a Grant from Science
Foundation Ireland under Grant number 17/CDA/4786. The work of K.-G
Nguyen was supported by HPY Research Foundation, Nokia Foundation,
Walter Ahlstr\"{o}m Foundation, Finnish Foundation for Technology
Promotion, Tauno T\"{o}nning Foundation.}\thanks{Kien-Giang Nguyen, Quang-Doanh Vu, and Markku Juntti are with Centre
for Wireless Communications, University of Oulu, FI-90014, Finland.
Email: \{giang.nguyen, doanh.vu, markku.juntti\}@oulu.fi.}\thanks{L.-N. Tran is with School of Electrical and Electronic Engineering,
University College Dublin, Ireland. Email: nam.tran@ucd.ie).}\vspace{-10mm}}
\maketitle
\begin{abstract}
We consider a multi-pair amplify-and-forward relay network where the
energy-constrained relays adopting time-switching protocol harvest
energy from the radio frequency signals transmitted by the users for
assisting user data transmission. Both one-way and two-way relaying
techniques are investigated. Aiming at energy efficiency (EE) fairness
among the user pairs, we construct an energy consumption model incorporating
rate-dependent signal processing power, the dependence on output power
level of power amplifiers' efficiency, and nonlinear energy harvesting
(EH) circuits. Then we formulate the max-min EE fairness problems
in which the data rates, users' transmit power, relays' processing
coefficient, and EH time are jointly optimized under the constraints
on the quality of service and users' maximum transmit power. To achieve
efficient suboptimal solutions to these nonconvex problems, we devise
monotonic descent algorithms based on the inner approximation (IA)
framework, which solve a second-order-cone program in each iteration.
To further simplify the designs, we propose an approach combining
IA and zero-forcing beamforming, which eliminates inter-pair interference
and reduces the numbers of variables and required iterations. Finally,
extensive numerical results are presented to validate the proposed
approaches. More specifically, the results demonstrate that ignoring
the realistic aspects of power consumption might degrade the performance
remarkably, and jointly designing parameters involved could significantly
enhance the energy efficiency.
\end{abstract}


\section{Introduction}

Relay-assisted cooperative communications can improve spectral and
energy efficiency, and, more importantly, extend the range of coverage
\cite{FengJiang:13:aSurveyofEE,tutorialAF_2012}. As such, relay-assisted
cooperative communications has been standardized in current mobile
networks, e.g.,\ 3GPP Long-Term Evolution (LTE) \cite{3gpprelease10overview}.
In addition, it is expected to be a major means to implement device-to-device
communications in the upcoming mobile networks \cite{D2DMag_2014}.
Various relay strategies have been proposed including amplify-and-forward
(AF), decode-and-forward (DF), and compress-and-forward \cite{surveyrelay_2012}.
Among them, AF has attracted significant interest due to its simplicity
and low latency \cite{tutorialAF_2012}.

Relaying can be either one-way or two-way. The former refers to one-directional
transmission from one network node to another, which is applied to
the scenario that only one node has data transmitted to another such
as the downlink transmission from an access point to a mobile phone.
The latter comprises a system in which both nodes send messages to
each other, introduced for improving spectral efficiency \cite{tutorialAF_2012}.
Two-way relaying was developed based on the self-interference cancellation
employed at the destinations to extract the desired signals \cite{Rankov2007JSAC}.

Early works on relay systems focused on single user pair, and for
improving spectral efficiency, a more general relay system including
multiple pairs of users was proposed \cite{Dehkordy2009TSP}. Here,
the relays simultaneously assist the transmission of multiple user
pairs forming an interference channel. Linear precoding at the relays
can be used to manage the radio resource and control the interference\cite{Joung2010TSP}.
In resource constrained networks such as wireless sensor networks,
the nodes are low-cost, i.e., each one is equipped with a single-antenna.
The benefits of MIMO techniques can be exploited if a pool of relays
collaboratively operate to perform the so called distributed relay
beamforming \cite{Ding2008WCOM}.

In a relay-based system where low-cost relays are equipped with limited
batteries, i.e.\ do not have sustainable power supplies, such as
sensors or mobile devices, one of the main implementation challenges
is to recharge the limited batteries for keeping the network alive
\cite{Huang2013JSAC}. To this end, simultaneous wireless information
and power transfer technique is a promising solution \cite{Huang2013JSAC,ZDing2014WCOM,Nasir2013WCOM,Lu2015EHSurveys}.
The technique allows the relays to harvest energy from the radio-frequency
(RF) signals, and thus the batteries can be wirelessly empowered.

Energy efficiency (EE) has become an important performance measure
in wireless networks \cite{ericssonEE}. By definition, the consumed
energy plays a vital role on EE objectives. Thus, the accuracy of
the power consumption model is crucial for designing practical systems.
For example, signal processing power and the efficiency of power amplifiers
(PAs) are commonly assumed to be fixed \cite{Zappone2014TSP,heliot2018TSP,Zhang2015SPL,Sheng2017WCOM}.
However, signal processing power is often rate-dependent \cite{Isheden:2010:globecom}
and PAs' efficiency depends on output power level \cite{AmplifierMIMO-Persson}.
It has been demonstrated that such aspects may have significant impacts
on the network level EE performances \cite{Osk2017TSP,OskSIP2016}.

\subsection*{Related Works}

Multi-pair one-way and two-way relaying have been investigated in
many prior works. \cite{Dehkordy2009TSP} considered a one-way relay
network with the aim of  minimizing the total transmit power at relays.
Therein, distributed relay beamforming was designed using the semidefinite
relaxation (SDR). This work was generalized in \cite{Cheng2012TSP}
where transmit power at users and the distributed relay beamforming
were designed for minimizing the total transmit power at users and
relays. The constrained concave convex procedure was used to tackle
the nonconvex problem. Similarly, in \cite{Wang2015WCOM}, the users'
power and relay beamforming were jointly designed for maximizing the
secrecy rate. On the other hand, \cite{Rankov2007JSAC} focused on
two-way relaying where the inter-pair interference is eliminated via
zero-forcing (ZF) relay beamforming. \cite{Joung2010TSP} considered
a system where a two-way relay is equipped with multiple antennas.
The processing matrix at the relay is designed based on ZF and minimum
mean-square-error criteria for achieving fairness among users and
maximizing system signal-to-noise ratio. \cite{Tao2012TSP} aimed
at achieving the max-min rate fairness among users. Therein, the relay's
processing matrix was designed by using the SDR and ZF. In general,
design problems for multiuser AF relay networks are nonconvex. Consequently,
the related works have mainly focused on suboptimal low-complexity
designs.

Cooperative systems with EH relays have received considerable attention.
In particular, \cite{Nasir2013WCOM} proposed time-switching and power-splitting
protocols for single user pair networks where a relay harvests energy
from the user's RF signal. To further improve the network performance,
the authors proposed dynamic EH time in \cite{Nasir2015WCOM}. A more
general system with multiple user pairs was considered in \cite{ZDing2014WCOM}.
Assuming user pairs use orthogonal channels, the work analyzed the
impacts of different power allocation strategies on the network performance.
\cite{Huang2013JSAC} considered a network where both users and the
relay harvest energy and focused on user and relay power allocation
for throughput maximization under the EH constraints. \cite{Huang2016WCOM}
considered multiple-input multiple-output (MIMO) AF system where a
relay simultaneously harvests energy transmitted from a destination
and receives information from a source. A system with a single user
pair and an EH two-way relay was studied in \cite{Chen2014Globalsip}.
A more general system with multiple EH relays and a single user pair
was recently studied in \cite{Liu2016COML} for one-way relaying,
and in \cite{salari2017WCOM} for two-way relaying. While \cite{Liu2016COML}
optimized the EH relays' power splitting ratio in order to maximize
the transmit data rate, the work in \cite{salari2017WCOM} jointly
designed EH time allocation and distributed relay beamforming for
three objectives including sum-rate maximization, total power consumption
minimization at relays, and EH time minimization.

EE for relay-assisted cooperative communications has recently been
studied. \cite{Zappone2014TSP} considered a one-way MIMO AF system
with one user pair and one relay. The work jointly optimized the user
and relay precoding matrices for different channel state information
assumptions. EE maximization for the similar system model, but with
a two-way relay, was studied in \cite{Zhang2015SPL}. More recently,
\cite{Sheng2017WCOM} solved the EE maximization for a two-way relay
network with multiple user pairs and multiple relays by jointly designing
user transmit power and relay matrices. \cite{Tan2017} considered
a multiple user pair one-way MIMO DF system. \cite{Zhang2017Access}
focused on a network with one user pair and one EH two-way relay,
and devised power allocation for maximizing EE performance. In the
aforementioned works, signal processing power and PAs' efficiency
were assumed to be constant. In a few recent publications \cite{Cui2017WCOM,cui2017VT},
the impacts of non-ideal PA efficiency and rate-dependent signal processing
power on the EE performance were studied for two-way systems with
one relay and one user pair. The EE problems for the network with
multiple user pairs and multiple EH relays have remained relatively
open in the literature.

\subsection*{Contributions}

Motivated by the above discussion and literature review, in this work,
we study the one-way and two-way multiuser AF relay networks where
the low-cost relays receive energy from the users for assisting data
transmission. The goal is to manage the EE fairness between the user
pairs, which is inspired from the fact that the users in a pair might
have to consume a lot of their own energy to charge the relays, while
the transmit data rate of the pair is small. Towards a relatively
realistic energy consumption model, we take into account the data-rate
signal processing power, the dependence of PAs' efficiency on the
output power level, and consider a practical model of EH circuit introduced
in \cite{Boshkovska2015COML}. Consequently, the parameters including
transmit data rate, users' transmit power, relays' processing coefficient,
and EH time, are mutually dependent, and should be jointly designed.
Hence, we formulate the problems of max-min EE fairness for both one-way
and two-way relay systems in which the mentioned parameters are optimization
variables.\footnote{We formulate the problems based on the EE definition, in which the
objective functions contain fractional functions. Another approach
for achieving EE in wireless communications is to minimize the power
consumption. However, as shown in many works (e.g.\ \cite{Doanh:15:SPL:EEforMultiHoming}),
EE performances obtained by this approach are far from optimal. } These problems inherit the numerical difficulties encountered in
multiuser AF relay networks, and thus, are nonconvex. We then develop
the low-complexity iterative algorithms based on the efficient descent
optimization framework, namely, inner approximation (IA) \cite{Beck2010,MarksWright:78:AGenInnerApprox}.\footnote{Another common suboptimal technique used for overcoming intractable
fractional EE problems is developed based on parametric fractional
programming, e.g.\ \cite{HHJY:13:JCOM}. However, this technique
may not be guaranteed to converge \cite[Section 4.1]{zappone2015energy}.} The convergence proofs for the algorithms are also provided. For
efficient practical implementations, we transform the convex approximate
problems into the second-order-cone programs (SOCPs), which is done
based on a concave lower bound of the logarithmic function. In addition,
for lower complexity designs, we develop solutions based on the combination
of IA and ZF beamforming which have smaller problem sizes, and thus
require fewer numbers of iterations to converge. Finally, we provide
extensive numerical results which confirm that our proposed approaches
are efficient in terms of the EE fairness. Specifically, the main
results indicate that realistic aspects of power consumption should
be taken into consideration in the EE designs, and much better performance
can be yielded by jointly optimizing parameters involved.

\emph{Organization: }The rest of the paper is organized as follows.
Section \ref{sec:System-Model-and} describes the system models and
formulates the problems. Section \ref{sec:Proposed-Algorithms} presents
the iterative algorithms developed based on IA. The designs based
on the combination of IA and ZF are provided in Section \ref{sec:Computational-Complexity}.
Section \ref{sec:Computational-Complexity-Analysi} discusses the
computational complexity of the proposed solutions. Numerical results
and discussion are provided in Section \ref{sec:Simulation-Result}.
Finally, Section \ref{sec:Conclusion} concludes the paper.

\emph{Notation:} Bold lower and upper case letters represent vectors
and matrices, respectively. $\left\Vert \cdot\right\Vert _{2}$ represents
the $\ell_{2}$ norm. $\left|\cdot\right|$ represents the absolute
value. $\mathbb{R}^{m\times n}$ and $\mathbb{C}^{m\times n}$ represent
the space of real and complex matrices of dimensions given in superscript,
respectively. $\mathbf{I}_{n}$ denotes the $n\times n$ identity
matrix. $\mathcal{CN}(0,c\mathbf{I})$ denotes a complex Gaussian
random vector with zero mean and variance matrix $c\mathbf{I}$. $\mathrm{\Re}(\cdot)$
represents real part of the argument. $\mathbf{A}\herm$ and $\mathbf{A}\trans$
are Hermitian and normal transpose of $\mathbf{A}$, respectively.
$\text{diag}({\bf a})$ represents diagonal matrix constructed from
element of ${\bf a}$. Notation $\odot$ stands for Schur-Hadamard
(element-wise) multiplication of two matrices. $\mathbf{e}_{l}\triangleq[\underset{l-1}{\underbrace{0,\ldots,0}},1,0,\ldots,0]$.
$[a]^{+}$denotes $\max(0,a)$. $\left\langle \mathbf{a},\mathbf{b}\right\rangle \triangleq\mathbf{a}\trans\mathbf{b}$
when $\mathbf{a}$ and $\mathbf{b}$ are real vectors, and $\left\langle \mathbf{a},\mathbf{b}\right\rangle \triangleq2\Re(\mathbf{a}\herm\mathbf{b})$
when $\mathbf{a}$ and $\mathbf{b}$ are complex vectors. Other notations
are defined at their first appearance.

\section{System Model and Problem Statement\label{sec:System-Model-and}}

In this section, we first describe the system model of multi-pair
relaying. Then the transmission protocol and energy consumption model
of one-way relaying are presented, following by those of two-way relaying.
Finally, the EE fairness problems are formulated.

\begin{figure}
\centering{}\includegraphics[width=0.45\columnwidth]{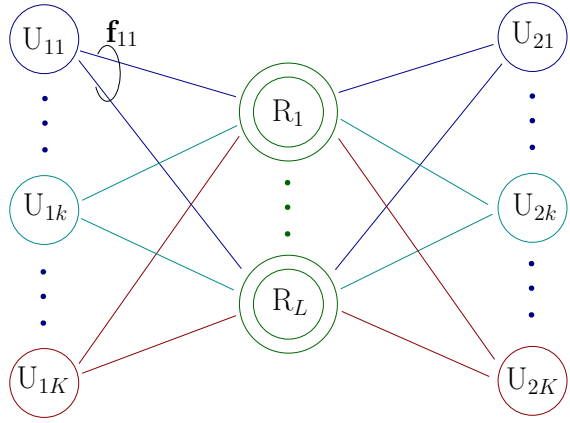}\caption{A diagram of multiple user pair AF relay systems with $K$ pairs of
users and $L$ relays.}
 \label{fig:systemrelayeh}
\end{figure}

We consider a multi-pair relay system consisting of a set of $K$
user pairs, denoted by ${\cal K}\triangleq\{1,\ldots,K\}$, and a
set of $L$ nonregenerative relays, denoted by ${\cal L}\triangleq\{1,\ldots,L\}$,
as shown in Figure \ref{fig:systemrelayeh}. Let us denote by $\mathrm{U}_{1k}$
and $\mathrm{U}_{2k}$ the two users of pair $k$,\footnote{Because we consider also two-way relaying, both nodes of each communicating
user pair play the role of source and destination. Therefore, we index
them as 1 and 2. In the one-way relay channel, 1 is the source and
2 is the destination, while in the two-way relaying both send and
receive.} and by $\mathrm{R}_{l}$ the relay $l$. Suppose that there is no
direct link between $\mathrm{U}_{1k}$ and $\mathrm{U}_{2k}$ for
any $k\in{\cal K}$, and a user intends to communicate within its
own pair with the help of the relays. All nodes operate in a half-duplex
mode and are low-cost, i.e., each of the nodes is equipped with a
single-antenna.

The channels are supposed to be flat block-fading with block time
$T$, and without loss of generality, let $T=1$ for notational simplicity.
Let $f_{ikl}$ denote the complex channel coefficient between $\mathrm{U}_{ik}$
and $\mathrm{R}_{l}$, and $\mathbf{f}_{ik}\triangleq[f_{ik1},...,f_{ikL}]\trans$.
The channel reciprocity holds for all links. Following \cite{Wang2015WCOM,Cheng2012TSP,Dehkordy2009TSP}
we suppose that perfect channel state information (CSI) is known at
a central node, where system optimization is performed.

We further assume that the transmit user nodes are non energy-constrained
while the relays are energy-constrained. Therefore, for assisting
the data transmission, the relays follow the time-switching protocol
to harvest energy from the RF signal transmitted from the users \cite{Nasir2013WCOM}.\footnote{Compared to power-splitting protocol, time-switching protocol requires
simpler hardware implementation (i.e.,\ simple switchers) \cite{Lu2015EHSurveys},
thus it is more suitable for low-cost nodes.} In particular, a transmission block is divided into two portions:
the first portion of duration $\tau$, $\tau\in(0,1)$, is a fraction
of block time used for charging the relays, referred to as EH phase.
The second portion is for the two-hop AF communications, referred
to as information transmission (IT) phase. In this work, we consider
both one-way and two-way relay systems. Communication protocol for
each of the systems is detailed below.

\subsection{One-Way Relay System\label{subsec:OW-Relay-Sys}}

In a one-way relay system, only one user in each pair transmits data
to the other. Without loss of generality and for notational convenience,
let us assume that $\mathrm{U}_{1k}$ is the transmitter and $\mathrm{U}_{2k}$
is the receiver, for all $k\in{\cal K}$.

\subsubsection{EH Phase (One-Way)}

During EH phase, the relays harvest energy from the RF signal transmitted
by the transmitters.\footnote{This scheme is for the scenario where it is inconvenient for the receivers
transmitting energy to the relays. For example, the receivers are
mobile phones with low batteries, and to transmit energy to the relays
could make the batteries run out quickly. } Particularly, the RF power at the input of the EH circuit of $\mathrm{R}_{l}$
is \cite{Nasir2013WCOM}
\begin{equation}
P_{l}^{\text{RF,OW}}({\bf p})\triangleq\sum_{k\in{\cal K}}p_{1k}|f_{1kl}|^{2}
\end{equation}
where $p_{ik}$,\textsf{ $(i=\{1,2\})$} is the transmit power at
$\mathrm{U}_{ik}$ and ${\bf p}\triangleq[p_{11},...,p_{1K}]\trans$.
The EH power circuit converts $P_{l}^{\text{RF,OW}}({\bf p})$ to
DC power used during the IT phase. Here, we consider a realistic RF-DC
power converter, whose conversion efficiency is not a constant, introduced
in \cite{Boshkovska2015COML}. Specifically, the harvested energy
at $\mathrm{R}_{l}$ is

\begin{equation}
E_{l}^{\textrm{EH,OW}}(\tau,{\bf p})=\frac{\tau\bar{P}_{l}^{\text{DC}}}{1-\beta_{l}}\left(\frac{1}{1+\exp\bigl(-c_{l}(P_{l}^{\text{RF,OW}}({\bf p})-d_{l})\bigr)}-\beta_{l}\right)\label{eq:RF-DC model}
\end{equation}
where $\bar{P}_{l}^{\text{DC}}$ is the maximum power that can be
harvested, $c_{l}$ and $d_{l}$ are parameters depending on the circuit
specifications, and $\beta_{l}=(1+\exp(c_{l}d_{l}))^{-1}$.

\subsubsection{IT Phase (One-Way)}

During IT phase, the remaining $(1-\tau)$ fraction of block time
is divided into two equal-length time slots. In the first time slot,
the transmitters send data to the relays. Let $x_{1k}$ denote the
normalized complex symbol transmitted by $\mathrm{U}_{1k}$. The
received signal at $\mathrm{R}_{l}$ is
\begin{equation}
\tilde{y}_{l}^{\textrm{OW}}=\sum_{k\in{\cal K}}\sqrt{p_{1k}}f_{1kl}x_{1k}+\tilde{n}_{l}
\end{equation}
where $\tilde{n}_{l}$ is the additive white Gaussian noise (AWGN),
i.e., $\tilde{{\bf n}}\sim\mathcal{CN}(0,\tilde{\sigma}^{2}\mathbf{I}_{L})$
with $\tilde{{\bf n}}\triangleq[\tilde{n}_{1},...,\tilde{n}_{L}]\trans$.
In the second time slot, the relays transmit the processed signal
to the receivers. We denote by $w_{l}\in\mathbb{C}$ the complex weight
coefficient used at $\mathrm{R}_{l}$, and let ${\bf w}\triangleq[w_{1},...,w_{L}]\trans\in\mathbb{C}^{L\times1}$.
The received signal at $\mathrm{U}_{2k}$ is
\begin{align}
y_{2k}^{\textrm{OW}} & =\sum_{l\in{\cal L}}f_{2kl}w_{l}\tilde{y}_{l}^{\text{OW}}+n_{2k}=\underset{\textrm{desired\,signal}}{\underbrace{\sqrt{p_{1k}}{\bf f}_{2k}\trans{\bf W}{\bf f}_{1k}x_{1k}}}+\underset{\textrm{interference}}{\underbrace{\sum_{j\in{\cal K}\backslash\{k\}}\sqrt{p_{1j}}{\bf f}_{2k}\trans{\bf W}{\bf f}_{1j}x_{1j}}}+\underset{\textrm{noise}}{\underbrace{{\bf f}_{2k}\trans{\bf W}\tilde{{\bf n}}+n_{2k}}}
\end{align}
where ${\bf W}\triangleq\text{diag}({\bf w})$, and $n_{2k}$ denotes
the additive noise with $n_{2k}\sim\mathcal{CN}(0,\sigma^{2})$. The
signal-to-interference-plus-noise ratio (SINR) at $\mathrm{U}_{2k}$
is
\begin{align}
\gamma_{2k}^{\textrm{OW}}({\bf w},{\bf p}) & =\frac{p_{1k}|{\bf f}_{2k}\trans{\bf W}{\bf f}_{1k}|^{2}}{\sum_{j\in{\cal K}\backslash\{k\}}p_{1j}|{\bf f}_{2k}\trans{\bf W}{\bf f}_{1j}|^{2}+\tilde{\sigma}^{2}||{\bf f}_{2k}\trans{\bf W}||_{2}^{2}+\sigma^{2}}\nonumber \\
 & =\frac{p_{1k}{\bf w}\herm{\bf H}_{kk}{\bf w}}{\sum_{j\in{\cal K}\backslash\{k\}}p_{1j}{\bf w}\herm{\bf H}_{kj}{\bf w}+{\bf w}\herm{\bf G}_{2k}{\bf w}+\sigma^{2}}
\end{align}
where $\mathbf{h}_{kj}\triangleq({\bf f}_{2k}\odot{\bf f}_{1j})\trans$,
${\bf H}_{kj}\triangleq\mathbf{h}_{kj}\herm\mathbf{h}_{kj}$, and
${\bf G}_{ik}\triangleq\tilde{\sigma}^{2}\text{diag}\bigl({\bf f}_{ik}\herm\odot{\bf f}_{ik}\trans)$.
Let $r_{ik}$ be the \emph{real} transmit data rate at $\mathrm{U}_{ik}$,
i.e., the effective information rate is $\frac{1-\tau}{2}r_{ik}$.
For feasible transmission, the constraint
\begin{equation}
r_{1k}\leq\log(1+\gamma_{2k}^{\textrm{OW}}({\bf w},{\bf p})),\,\forall k\in{\cal K}\label{eq:rate}
\end{equation}
should hold. The purpose of introducing $\{r_{1k}\}_{k=1}^{K}$ is
to determine the rate-dependent signal processing energy, which is
discussed in detail next.

\subsubsection{Energy Consumption Model (One-Way)}

We consider herein a relatively realistic energy consumption model
which takes into account the dependence of PAs' efficiency on the
output power level \cite{Mikami2007,AmplifierMIMO-Persson} as well
as the dependence of signal processing operators on the transmit data
rate \cite{Isheden:2010:globecom}. In addition, for saving energy,
a node can be idle (i.e.,\ sleep mode) if it is neither receiving
nor transmitting \cite{heliot2018TSP,Cui2017WCOM}. In this spirit,
let us first focus on the energy consumed by the users of pair $k$.
Let $P_{ik}^{\textrm{idle}}$ denote the consumed power of $\mathrm{U}_{ik}$
in idle mode, which is assumed to be constant \cite{heliot2018TSP,Cui2017WCOM}.
Then, the energy consumed in this mode is
\begin{equation}
E_{k}^{\text{idle}}(\tau)=\frac{1-\tau}{2}P_{1k}^{\textrm{idle}}+\frac{1+\tau}{2}P_{2k}^{\textrm{idle}}.
\end{equation}
On the other hand, for the clarity of description, we divide the power
consumed in the active mode into three components: power consumed
by the operating circuits, the amplifiers and signal processing. The
first part includes the power consumed, e.g., by filters, mixers,
etc, denoted by $P_{ik}^{\textrm{act,cir}}$ for $\mathrm{U}_{ik}$.
It is modeled as a constant \cite{HowmuchEnergy-Auer}. For the power
consumed on the amplifiers, we consider a realistic model whose efficiency
is given by \cite[Eq. (2)]{AmplifierMIMO-Persson}
\begin{equation}
\tilde{\epsilon}_{ik}=\epsilon_{ik}\sqrt{\frac{p_{ik}}{\bar{P}_{ik}}}\label{eq:ampeff}
\end{equation}
where $\epsilon_{ik}\in(0,1)$ is the maximum PA's efficiency and
$\bar{P}_{ik}$ is the maximum transmit power of $\mathrm{U}_{ik}$.
From (\ref{eq:ampeff}), the power consumed on the PA is
\begin{equation}
P_{ik}^{\textrm{amp}}=\frac{p_{ik}}{\tilde{\epsilon}_{ik}}=\varepsilon_{ik}\sqrt{p_{ik}}\label{eq:realpowamp}
\end{equation}
 where $\varepsilon_{ik}=\sqrt{\bar{P}_{ik}}/\epsilon_{ik}$. Finally,
the power for signal processing is modeled as a linear function of
data rate given by $P_{1k}^{\textrm{sp}}=(\rho_{1k}^{\textrm{en}}+\rho_{2k}^{\textrm{de}})r_{1k}$
where $\rho_{1k}^{\textrm{en}}$ and $\rho_{2k}^{\textrm{de}}$ represent
power for encoder at $\mathrm{U}_{1k}$ and decoder at $\mathrm{U}_{2k}$,
respectively. Their units are in W/(Gnats/s). In summary, the total
energy consumed by pair $k$ during a block time is
\begin{align}
E_{k}^{\textrm{OW}}(\tau,{\bf p},{\bf r}) & =E_{k}^{\text{idle}}(\tau)+\frac{1+\tau}{2}(P_{1k}^{\textrm{amp}}+P_{1k}^{\textrm{act,cir}})+\frac{1-\tau}{2}(P_{1k}^{\textrm{sp}}+P_{2k}^{\textrm{act,cir}})\nonumber \\
 & =\frac{1-\tau}{2}(\rho_{1k}^{\textrm{sp}}r_{1k}+P'_{k})+\frac{1+\tau}{2}(\varepsilon_{1k}\sqrt{p_{1k}}+P''_{k})\label{eq:power:consump:link}
\end{align}
where $\rho_{1k}^{\textrm{sp}}=\rho_{1k}^{\textrm{en}}+\rho_{2k}^{\textrm{de}}$,
$P'_{k}=P_{1k}^{\textrm{idle}}+P_{2k}^{\textrm{act,cir}}$, and $P''_{k}=P_{2k}^{\textrm{idle}}+P_{1k}^{\textrm{act,cir}}$,
which are constant; ${\bf r}\triangleq[r_{11},...,r_{1K}]\trans$.

We now describe the energy consumed by the relays. The radiated power
at $\mathrm{R}_{l}$ is $P_{l}^{\text{rad}}({\bf p},{\bf w})\triangleq w_{l}^{\ast}\Bigl({\textstyle \sum_{k\in{\cal K}}}p_{1k}|f_{1kl}|^{2}+\tilde{\sigma}^{2}\Bigr)w_{l}={\bf w}\herm{\bf A}_{l}{\bf w}+\sum_{k\in{\cal K}}p_{1k}{\bf w}\herm{\bf B}_{1kl}{\bf w}$
where ${\bf A}_{l}\triangleq\tilde{\sigma}^{2}\text{diag}(\mathbf{e}_{l})$
and ${\bf B}_{1kl}\triangleq|f_{1kl}|^{2}\text{diag}(\mathbf{e}_{l})$.
Then, the total energy consumed at $\mathrm{R}_{l}$ is given by
\begin{equation}
E_{l}^{\textrm{R,OW}}(\tau,{\bf p},{\bf w})\triangleq\frac{1-\tau}{2}\frac{\sqrt{\bar{P}_{l}^{\text{R}}P_{l}^{\text{rad}}({\bf p},{\bf w})}}{\epsilon_{l}^{\text{R}}}+E_{l}^{\textrm{R,const}}\label{eq:power:consume:relay}
\end{equation}
where $\epsilon_{l}^{\text{R}}\in(0,1)$ is the maximum PA's efficiency
and $\bar{P}_{l}^{\text{R}}$ is the maximum transmit power of $\mathrm{R}_{l}$.
In (\ref{eq:power:consume:relay}), the first term is the energy consumed
by the PA, and $E_{l}^{\textrm{R,const}}$ is the consumed energy
for activating the basic functions which is constant \cite{HowmuchEnergy-Auer}.
Since the relays do not encode or decode data, the rate-dependent
signal processing energy does not exist. Clearly, for successfully
assisting the data transmission, the energy consumption cannot exceed
harvesting or
\begin{equation}
E_{l}^{\text{R,OW}}(\tau,{\bf p},{\bf w})\leq E_{l}^{\textrm{EH,OW}}(\tau,{\bf p}),\,\forall l\in{\cal L}.\label{eq:EnergyCons}
\end{equation}

\subsection{Two-Way Relay System}

In a two-way system, the relays assist the bi-directional communication
of all pairs, i.e., both of the two users of each pair transmit and
receive data.

\subsubsection{EH Phase (Two-Way)}

The relays receive energy from the both two users of each pair. Hence,
the RF power at the input of EH circuit of $\mathrm{R}_{l}$ is
\begin{equation}
P_{l}^{\text{RF,TW}}(\tilde{\mathbf{p}})\triangleq\sum_{k\in{\cal K}}\sum_{i=1}^{2}p_{ik}|f_{ikl}|^{2}
\end{equation}
where $\tilde{\mathbf{p}}\triangleq[p_{11},p_{21},...,p_{1K},p_{2K}]\trans$.
Accordingly, the harvested energy at $\mathrm{R}_{l}$ is

\begin{equation}
E_{l}^{\textrm{EH,TW}}(\tau,\tilde{\mathbf{p}})=\frac{\tau\bar{P}_{l}^{\text{DC}}}{1-\beta_{l}}\left(\frac{1}{1+\exp\bigl(-c_{l}(P_{l}^{\text{RF,TW}}(\tilde{\mathbf{p}})-d_{l})\bigr)}-\beta_{l}\right).\label{eq:RF-DC model-1}
\end{equation}

\subsubsection{IT Phase (Two-Way)}

In the first time slot of IT phase, all the users transmit their signals
to the relays using the same frequency band. Particularly, the received
signal at $\mathrm{R}_{l}$ is
\begin{equation}
\tilde{y}_{l}^{\textrm{TW}}=\sum_{k\in{\cal K}}\sum_{i=1}^{2}\sqrt{p_{ik}}f_{ikl}x_{ik}+\tilde{n}_{l}.
\end{equation}
During the second time slot, the relays broadcast the processed signals
to all the users. The received signal at $\mathrm{U}_{ik}$ is expressed
as
\begin{align}
y_{ik}^{\textrm{rec}} & =\sum_{l\in{\cal L}}f_{ikl}w_{l}\tilde{y}_{l}^{\textrm{TW}}+n_{ik}=\sum_{j\in{\cal K}}\sum_{\hat{i}=1}^{2}\sqrt{p_{\hat{i}j}}{\bf f}_{ik}\trans{\bf W}{\bf f}_{\hat{i}j}x_{\hat{i}j}+{\bf f}_{ik}\trans{\bf W}\tilde{{\bf n}}+n_{ik}.
\end{align}
As with most of the related works (see \cite{Rankov2007JSAC,Joung2010TSP,Tao2012TSP,Sheng2017WCOM}
and the references therein), we suppose that the self-interference
can be completely canceled at the users (with the known CSI). Then
the signal for decoding at $\mathrm{U}_{ik}$ reduces to
\begin{equation}
y_{ik}^{\textrm{TW}}=\underset{\textrm{desired\,signal}}{\underbrace{\sqrt{p_{\bar{i}k}}{\bf f}_{ik}\trans{\bf W}{\bf f}_{\bar{i}k}x_{\bar{i}k}}}+\underset{\textrm{interference}}{\underbrace{\sum_{j\in{\cal K}\setminus\{k\}}\sum_{\hat{i}=1}^{2}\sqrt{p_{\hat{i}j}}{\bf f}_{ik}\trans{\bf W}{\bf f}_{\hat{i}j}x_{\hat{i}j}}}+\underset{\textrm{noise}}{\underbrace{{\bf f}_{ik}\trans{\bf W}\tilde{{\bf n}}+n_{ik}}}
\end{equation}
where $\bar{i}=\{1,2\}\setminus\{i\}$. Thus the SINR at $\mathrm{U}_{ik}$
can be written as
\begin{align}
\gamma_{ik}^{\textrm{TW}}({\bf w},\tilde{\mathbf{p}}) & =\frac{p_{\bar{i}k}|{\bf f}_{ik}\trans{\bf W}{\bf f}_{\bar{i}k}|^{2}}{\sum_{j\in{\cal K}\setminus\{k\}}\sum_{\hat{i}=1}^{2}p_{\hat{i}j}|{\bf f}_{ik}\trans{\bf W}{\bf f}_{\hat{i}j}|^{2}+\tilde{\sigma}^{2}||{\bf f}_{ik}\trans{\bf W}||_{2}^{2}+\sigma^{2}}\nonumber \\
 & =\frac{p_{\bar{i}k}{\bf w}\herm{\bf H}_{kk}{\bf w}}{\sum_{j\in{\cal K}\backslash\{k\}}\sum_{\hat{i}=1}^{2}p_{\hat{i}j}{\bf w}\herm\tilde{{\bf H}}_{ik\hat{i}j}{\bf w}+{\bf w}\herm{\bf G}_{ik}{\bf w}+\sigma^{2}}
\end{align}
where $\tilde{{\bf h}}_{ik\hat{i}j}\triangleq({\bf f}_{ik}\odot{\bf f}_{\hat{i}j})\trans$
and $\tilde{{\bf H}}_{ik\hat{i}j}\triangleq\tilde{{\bf h}}_{ik\hat{i}j}\herm\tilde{{\bf h}}_{ik\hat{i}j}$.
We note that $\tilde{{\bf H}}_{ik\bar{i}k}=\tilde{{\bf H}}_{\bar{i}kik}={\bf H}_{kk}$.
Similar to the one-way system, we need the following set of constraints
for successful transmissions
\begin{equation}
r_{ik}\leq\log(1+\gamma_{\bar{i}k}^{\textrm{TW}}({\bf w},\tilde{\mathbf{p}})),\forall k\in{\cal K},i=\{1,2\}.\label{eq:rate-tw}
\end{equation}

\subsubsection{Energy Consumption Model (Two-Way)}

Different from the one-way relay system, the users in the two-way
relay system are always active since each of them either transmits
or receives during block time. In addition, the energy for the power
amplifiers accounts on the both users of a pair, and the rate-dependent
signal processing energy for pair $k$ is calculated based on the
rate transmitted from $\mathrm{U}_{1k}$ and $\mathrm{U}_{2k}$. Thus
the energy consumed by pair $k$ can be expressed as
\begin{align}
E_{k}^{\textrm{TW}}(\tau,\tilde{\mathbf{p}},\tilde{\mathbf{r}}) & \triangleq\underset{\textrm{energy\,for\,circuits}}{\underbrace{1\left(\sum_{i=1}^{2}P_{ik}^{\textrm{act,cir}}\right)}}+\underset{\textrm{energy\,for\,PAs}}{\underbrace{\frac{1+\tau}{2}\left(\sum_{i=1}^{2}P_{ik}^{\textrm{amp}}\right)}}+\underset{\textrm{energy\,for\,signal\,processing}}{\underbrace{\frac{1-\tau}{2}\left(\sum_{i=1}^{2}P_{ik}^{\text{sp}}\right)}}\nonumber \\
 & =E_{k}^{\textrm{TW,cir}}+\frac{1+\tau}{2}\left(\sum_{i=1}^{2}\varepsilon_{ik}\sqrt{p_{ik}}\right)+\frac{1-\tau}{2}\left(\sum_{i=1}^{2}\rho_{ik}^{\textrm{sp}}r_{ik}\right)
\end{align}
where $E_{k}^{\textrm{TW,cir}}=1(\sum_{i=1}^{2}P_{ik}^{\textrm{act,cir}})$
is a constant, $\rho_{ik}^{\textrm{sp}}=\rho_{ik}^{\textrm{en}}+\rho_{\bar{i}k}^{\textrm{de}}$,
and $\tilde{\mathbf{r}}\triangleq[r_{11},r_{21},...,r_{1K},r_{2K}]\trans$.

For two-way relay $\mathrm{R}_{l}$, the radiated power is $P_{l}^{\text{rad,TW}}(\tilde{\mathbf{p}},{\bf w})\triangleq w_{l}^{\ast}\Bigl({\textstyle \sum_{k\in{\cal K}}}\sum_{i=1}^{2}p_{ik}|f_{ikl}|^{2}+\tilde{\sigma}^{2}\Bigr)w_{l}={\bf w}\herm{\bf A}_{l}{\bf w}+\sum_{k\in{\cal K}}\sum_{i=1}^{2}p_{ik}{\bf w}\herm{\bf B}_{ikl}{\bf w}$.
Then, the total consumed energy at $\mathrm{R}_{l}$ is
\begin{equation}
E_{l}^{\text{R,TW}}(\tau,\tilde{\mathbf{p}},{\bf w})\triangleq\frac{1-\tau}{2}\frac{\sqrt{\bar{P}_{l}^{\text{R}}P_{l}^{\text{rad,TW}}({\bf w},\tilde{\mathbf{p}})}}{\epsilon_{l}^{\text{R}}}+E_{l}^{\textrm{R,const}}.\label{eq:power:consume:relay-1}
\end{equation}
Again, the following set of constraints on the harvested and consumed
energy is required for successful relaying
\begin{equation}
E_{l}^{\text{R,TW}}(\tau,\tilde{\mathbf{p}},{\bf w})\leq E_{l}^{\textrm{EH,TW}}(\tau,\tilde{\mathbf{p}}),\,\forall l\in{\cal L}.\label{eq:EnergyCons-tw}
\end{equation}

\subsection{Energy Efficiency Fairness Problems}

We focus on the max-min EE. Here, the shared relays use energy contributed
by the users for assisting data transmission, when each user exchanges
information with the one in the same pair only. Hence, it is relevant
to maintain the EE fairness (EEF) between the user pairs.

\subsubsection{EEF for One-Way Relay System}

With the model specified in Section \ref{subsec:OW-Relay-Sys} and
by definition, the individual EE of pair $k$ is given by
\begin{equation}
f_{k}^{\text{EE,OW}}(\tau,{\bf p},{\bf r})\triangleq\frac{\frac{1-\tau}{2}r_{1k}}{E_{k}^{\text{OW}}(\tau,{\bf p},{\bf r})},\ k\in{\cal K}
\end{equation}
Thereby the problem of max-min EEF can be mathematically formulated
as \begin{subequations}\label{Prob:general}
\begin{align}
\underset{{\bf p},{\bf w},{\bf r},\tau}{\text{maximize}}\  & \min_{1\leq k\leq K}\ f_{k}^{\text{EE,OW}}(\tau,{\bf p},{\bf r})\label{eq:obj}\\
\text{subject to}\  & \frac{1-\tau}{2}r_{1k}\ge Q_{1k},\,\forall k\in{\cal K}\label{eq:QoS}\\
\textrm{EEF-OW}\triangleq\smash{\left\{ \IEEEstrut[13\jot]\right.}\hspace{1.8cm} & 0<p_{1k}\leq\bar{P}_{1k},\,\forall k\in{\cal K}\label{eq:source:txpower}\\
\; & P_{l}^{\text{rad}}({\bf p},{\bf w})\leq\bar{P}_{l}^{\text{R}},\,\forall l\in{\cal L}\label{eq:relaypow}\\
\; & \eqref{eq:rate},\eqref{eq:EnergyCons}.\label{eq:relaypower}
\end{align}
\end{subequations}Constraint (\ref{eq:QoS}) guarantees the quality
of service (QoS) for each user pair, where $Q_{1k}>0$ is a predefined
threshold. (\ref{eq:source:txpower}) and (\ref{eq:relaypow}) represent
the transmit power constraints at the transmitters and the relays,
respectively.

\subsubsection{EEF for Two-Way Relay System}

Similarly, we obtain the problem of max-min EEF for the two-way system
as\begin{subequations}\label{Prob:general-tw}\vspace{-3mm}
\begin{align}
\underset{\tilde{\mathbf{p}},{\bf w},\tilde{\mathbf{r}},\tau}{\text{maximize}}\  & \min_{1\leq k\leq K}\ f_{k}^{\text{EE,TW}}(\tau,\tilde{\mathbf{p}},\tilde{\mathbf{r}})\triangleq\frac{\frac{1-\tau}{2}\sum_{i=1}^{2}r_{ik}}{E_{k}^{\textrm{TW}}(\tau,\tilde{\mathbf{p}},\tilde{\mathbf{r}})}\label{eq:obj-tw}\\
\text{subject to}\  & \frac{1-\tau}{2}r_{ik}\ge Q_{ik},\,\forall k\in{\cal K},i=1,2\label{eq:QoS-tw}\\
\textrm{EEF-TW}\triangleq\smash{\left\{ \IEEEstrut[13\jot]\right.}\hspace{1.8cm} & 0<p_{ik}\leq\bar{P}_{ik},\,\forall k\in{\cal K},i=1,2\label{eq:source:txpower-tw}\\
\; & P_{l}^{\text{rad,TW}}(\tilde{\mathbf{p}},{\bf w})\leq\bar{P}_{l}^{\text{R}},\,\forall l\in{\cal L}\label{eq:relaypow-tw}\\
\; & \eqref{eq:rate-tw},\eqref{eq:EnergyCons-tw}.\label{eq:relaypower-tw}
\end{align}
\end{subequations}

In this work, we assume that the feasible sets of EEF-OW and EEF-TW
are nonempty. The objectives in EEF-OW and EEF-TW are nonsmooth nonconvex\textemdash the
numerators of the fractions are linear, but the denominators are nonconvex.
Also, the feasible sets are nonconvex. Hence the problems are intractable
and it is impossible to transform the problems into the equivalent
convex ones. Like many studies on wireless communication designs \cite{salari2017WCOM,Sheng2017WCOM,Wang2015WCOM,Huang2016WCOM},
 we aim at finding approximate, but efficient, solutions to these
problems. \vspace{-3mm}

\section{The Proposed Algorithms for Solving EEF-OW and EEF-TW\label{sec:Proposed-Algorithms}}

In this section, we propose algorithms for solving EEF-OW and EEF-TW
based on the inner approximation (IA) framework \cite{Beck2010,MarksWright:78:AGenInnerApprox},
which is an efficient approach widely used for dealing with nonconvex
programs. First, the general principles of the IA and the useful approximation
functions are provided. Then, the IA-based algorithms solving EEF-OW
and EEF-TW are presented, followed by the convergence discussion.
Finally, the approach arriving at the SOCP approximations is provided.
\vspace{-3mm}

\subsection{Useful Approximate Formulations\label{subsec:AppForm}}

For exposition purpose, we first provide some approximate formulations
which are used to devise proposed solutions. Generally the basic idea
of IA is to successively approximate a nonconvex set to inner convex
ones. Specifically, let $h(\mathbf{x})\leq0$ be a nonconvex constraint
where $h(\mathbf{x}):\mathbb{C}^{n}\rightarrow\mathbb{R}$ and $h(\mathbf{x})$
is continuously differentiable. An \emph{inner} approximation is obtained
by replacing $h(\mathbf{x})$ by a convex upper bound $\tilde{h}(\mathbf{x};g(\mathbf{x}'))$,
i.e., $h(\mathbf{x})\leq\tilde{h}(\mathbf{x};g(\mathbf{x}'))$, where
$g(\mathbf{x}):\mathbb{C}^{n}\rightarrow\mathbb{C}^{m}$ is a \emph{parameter
vector} and $\mathbf{x}'$ is some feasible point. Function $\tilde{h}(\mathbf{x};g(\mathbf{x}'))$
must satisfy the following conditions
\begin{equation}
h(\mathbf{x})=\tilde{h}(\mathbf{x};g(\mathbf{x})),\,\nabla_{\mathbf{x}^{\ast}}h(\mathbf{x})=\nabla_{\mathbf{x}^{\ast}}\tilde{h}(\mathbf{x};g(\mathbf{x}))\label{eq:IAcondition}
\end{equation}
where $\nabla_{\mathbf{x}^{\ast}}h()$ denotes the gradient of $h()$
with respect to the complex conjugate of $\mathbf{x}$. If $\mathbf{x}$
is a real vector, then $\nabla_{\mathbf{x}^{\ast}}h()$ is simply
replaced by $\nabla_{\mathbf{x}}h()$. The approximations presented
next follow these principles.

\subsubsection{Approximation for Bilinear Function}

Consider nonconvex constraint $x_{1}x_{2}\leq y$ where $(x_{1},x_{2},y)\in\mathbb{R}_{++}^{3}$.
An approximation of bilinear function $x_{1}x_{2}$ is given by \cite[Lem. 3.5]{Beck2010}
\begin{equation}
x_{1}x_{2}\leq h^{\textrm{bi}}(x_{1},x_{2};\lambda)\triangleq0.5\left(\lambda x_{1}^{2}+\frac{x_{2}^{2}}{\lambda}\right)\label{eq:Biappr}
\end{equation}
where $\lambda=\frac{x'_{2}}{x'_{1}}$. We remark that the bilinear
function can be rewritten as difference-of-convex ones, e.g.,\ $x_{1}x_{2}=0.25(x_{1}+x_{2})^{2}-0.25(x_{1}-x_{2})^{2}=0.5(x_{1}+x_{2})^{2}-0.5(x_{1}^{2}+x_{2}^{2})=0.5(x_{1}^{2}+x_{2}^{2})-0.5(x_{1}-x_{2})^{2}$.
Then the approximates can be obtained by using the first order Taylor
series approximation of the nonconvex parts. Herein, we use (\ref{eq:Biappr})
for problems EEF-OW and EEF-TW, since we numerically observe that
with (\ref{eq:Biappr}), the iterative procedures require fewer number
of iterations for convergence (see Fig. \ref{Fig.3.a} for the numerical
example).

\subsubsection{Approximation for Fractional-Linear Function}

Consider nonconvex constraint $\frac{x_{1}}{x_{2}}\leq y$ where $(x_{1},x_{2},y)\in\mathbb{R}_{++}^{3}$.
In light of (\ref{eq:Biappr}), an approximation of fractional-linear
function $\frac{x_{1}}{x_{2}}$ can be obtained as
\begin{equation}
\frac{x_{1}}{x_{2}}\leq h^{\textrm{frac}}(x_{1},x_{2};\lambda)\triangleq0.5\left(\lambda x_{1}^{2}+\frac{1}{\lambda x_{2}^{2}}\right)\label{eq:Biappr-1}
\end{equation}
where $\lambda=\frac{1}{x'_{1}x'_{2}}$. We note that constraint $h^{\textrm{frac}}(x_{1},x_{2};\lambda)\leq y$
can be expressed by the following two second-order cone (SOC) ones
\begin{equation}
0.5\left(\lambda x_{1}^{2}+\frac{z^{2}}{\lambda}\right)\leq y,\,1\leq x_{2}z.
\end{equation}

\subsubsection{Approximation for Quadratic-over-Linear Function}

Consider concave function $h(\mathbf{x},z;\mathbf{A})\triangleq-\frac{\mathbf{x}\herm\mathbf{A}\mathbf{x}}{z}$
where $\mathbf{x}\in\mathbb{C}^{n}$, $z\in\mathbb{R}_{++}$, and
$\mathbf{A}\succeq0$. We can use the first order Taylor series to
obtain a convex upper bound of $h(\mathbf{x},z;\mathbf{A})$ given
as
\begin{align}
h^{\textrm{qol}}(\mathbf{x},z;\mathbf{x}',z';\mathbf{A}) & \triangleq h(\mathbf{x}',z';\mathbf{A})+\left\langle [\nabla_{\mathbf{x}^{\ast}}h(\mathbf{x}',z';\mathbf{A}),\nabla_{z}h(\mathbf{x}',z';\mathbf{A})]\trans,[\mathbf{x}-\mathbf{x}',z-z']\trans\right\rangle \nonumber \\
 & =\frac{(\mathbf{x}')\herm\mathbf{A}\mathbf{x}'}{(z')^{2}}z-\frac{2\Re(({\bf x}')\herm\mathbf{A}{\bf x})}{z'}.
\end{align}

\subsubsection{Approximation for Logarithmic Function}

Consider logarithmic function $h(x)\triangleq\log(x)$ where $x\in\mathbb{R}_{++}$.
An approximated function of $h(x)$ is given by
\begin{equation}
h(x)\leq h^{\textrm{log}}(x;x')\triangleq\log(x')-1+\frac{x}{x'}.
\end{equation}

\subsubsection{Approximation for Power Function}

Consider power function $h(x;m)\triangleq-x^{m}$ where $x\in\mathbb{R}_{++}$.
Here, we only focus on the cases $m<0$ or $m>1$ where $h(x;m)$
is concave. Its convex approximation is given by
\begin{equation}
h(x;m)\leq h^{\textrm{po}}(x;x';m)\triangleq(m-1)(x')^{m}-m(x')^{m-1}x.
\end{equation}

\subsection{Solution for EEF-OW}

Directly applying IA to (\ref{Prob:general}) is difficult, since
the nonconvex parts here are not explicitly exposed. As a necessary
step, we translate (\ref{Prob:general}) into an equivalent, but more
tractable, formulation. We first introduce variable $\eta>0$ and
arrive at the epigraph form of (\ref{Prob:general}) given as\begin{subequations}\label{Prob:epiprob}
\begin{align}
\underset{{\bf p},{\bf w},{\bf r},\tau,\eta}{\text{minimize}}\  & \eta\label{eq:obj-1}\\
\text{subject to}\  & f_{k}^{\text{EE,OW}}(\tau,{\bf p},{\bf r})\geq\eta^{-1},\,\forall k\in{\cal K}\label{eq:eeconst}\\
\; & \eqref{eq:rate},\eqref{eq:EnergyCons},\eqref{eq:QoS},\eqref{eq:source:txpower},\eqref{eq:relaypow}.\label{eq:constraints}
\end{align}
\end{subequations}Here the nonconvex parts include (\ref{eq:rate}),
(\ref{eq:EnergyCons}), (\ref{eq:relaypow}), and (\ref{eq:eeconst}).

\subsubsection{Changes of Variables\label{subsec:Changes-of-Variables}}

We now make some changes of variables. Specifically, we will denote
$q_{ik}=\frac{1}{p_{ik}}$, $\forall k\in{\cal K},i=1,2$, and turn
the nonconvex products of linear and quadratic functions, e.g., $p_{1k}{\bf w}\herm{\bf H}_{kk}{\bf w}$,
into the quadratic-over-linear functions. We also define $\tilde{\tau}=\frac{1+\tau}{1-\tau}$,
i.e.,\ $\tau=\frac{\tilde{\tau}-1}{\tilde{\tau}+1}$. It is important
to note that these changes of variables still preserve the convexity
in (\ref{eq:QoS}) and (\ref{eq:source:txpower}) as well as turn
nonconvex constraint (\ref{eq:relaypow}) into a convex one. In addition,
they make (\ref{eq:rate}), (\ref{eq:EnergyCons}), and (\ref{eq:eeconst})
become more convenient to handle, as shown next.

\subsubsection{Transformation of (\ref{eq:rate})}

By introducing new variables $\{v_{k}\}_{k=1}^{K}$ and $\{s_{k}\}_{k=1}^{K}$,
we can equivalently represent (\ref{eq:rate}) by the following set
of constraints
\begin{gather}
r_{1k}\leq\log(1+v_{k}),\,\forall k\in{\cal K}\label{eq:rate-1}\\
\sum_{j\in{\cal K}\backslash\{k\}}\frac{{\bf w}\herm{\bf H}_{kj}{\bf w}}{q_{1j}}+{\bf w}\herm{\bf G}_{2k}{\bf w}+\sigma^{2}\leq s_{k},\,\forall k\in{\cal K}\label{eq:rate-2}\\
s_{k}v_{k}\leq\frac{{\bf w}\herm{\bf H}_{kk}{\bf w}}{q_{1k}},\,\forall k\in{\cal K}.\label{eq:rate-3}
\end{gather}
Here, only (\ref{eq:rate-3}) is nonconvex which contains bilinear
and quadratic-over-linear functions.

\subsubsection{Transformation of (\ref{eq:EnergyCons})}

We first rewrite (\ref{eq:EnergyCons}) with the change variables
as
\begin{align*}
\tilde{\varepsilon}_{l}\sqrt{{\bf w}\herm{\bf A}_{l}{\bf w}+\sum_{k\in{\cal K}}\frac{{\bf w}\herm{\bf B}_{1kl}{\bf w}}{q_{1k}}} & \leq\frac{\hat{\beta}_{l}(\tilde{\tau}-1)}{1+\alpha_{l}\exp\bigl(-c_{l}\sum_{k\in{\cal K}}\frac{|f_{1kl}|^{2}}{q_{1k}}\bigr)}-\bar{\beta}_{l}\tilde{\tau}+\check{\beta}_{l},\!\forall l\in\mathcal{L}
\end{align*}
where $\tilde{\varepsilon}_{l}=\sqrt{\bar{P}_{l}^{\text{R}}}/\epsilon_{l}^{\text{R}}$,
$\hat{\beta}_{l}=\frac{\bar{P}_{l}^{\text{DC}}}{1-\beta_{l}}$, $\bar{\beta}_{l}=\beta_{l}\hat{\beta}_{l}+E_{l}^{\textrm{R,const}}$,
$\check{\beta}_{l}=\bar{\beta}_{l}-2E_{l}^{\textrm{R,const}}$ and
$\alpha_{l}=\exp(c_{l}d_{l})$. Also, to reveal the hidden convexity
in the constraint, we introduce new variables $\{u_{l}\}_{l=1}^{L}$
and $\{t_{l}\}_{l=1}^{L}$, and equivalently rewrite (\ref{eq:EnergyCons})
as
\begin{gather}
\sum_{k\in{\cal K}}\frac{{\bf w}\herm{\bf B}_{1kl}{\bf w}}{q_{1k}}\leq u_{l}^{2},\!\forall l\in\mathcal{L}\label{eq:Econst-1}\\
\log(\tilde{\tau}-t_{l}-1)-\log(\alpha_{l}t_{l})+\sum_{k\in{\cal K}}\frac{c_{l}|f_{1kl}|^{2}}{q_{1k}}\geq0,\!\forall l\in\mathcal{L}\label{eq:Econst-2}\\
\tilde{\varepsilon}_{l}\sqrt{{\bf w}\herm{\bf A}_{l}{\bf w}+u_{l}^{2}}\leq\hat{\beta}_{l}t_{l}-\bar{\beta}_{l}\tilde{\tau}+\check{\beta}_{l},\!\forall l\in\mathcal{L}.\label{eq:Econst-3}
\end{gather}
The nonconvex parts are in (\ref{eq:Econst-1}) and (\ref{eq:Econst-2})
including the power and the logarithmic functions.

\subsubsection{Transformation of (\ref{eq:eeconst})}

Constraint (\ref{eq:eeconst}) is rewritten as
\begin{equation}
\rho_{1k}^{\text{sp}}+\frac{P'_{k}}{r_{1k}}+\frac{\tilde{\tau}}{r_{1k}}\Bigl(\frac{\varepsilon_{1k}}{\sqrt{q_{1k}}}+P''_{k}\Bigr)\leq\eta,\!\forall k\in\mathcal{K},\label{eq:eesource-equi}
\end{equation}
which is equivalently represented as
\begin{gather}
\frac{\tilde{\tau}}{r_{1k}}\leq z_{k}^{2},\!\forall k\in\mathcal{K}\label{eq:eesource-1}\\
\rho_{1k}^{\text{sp}}+\frac{P'_{k}}{r_{1k}}+\varepsilon_{1k}\frac{z_{k}^{2}}{\sqrt{q_{1k}}}+P''_{k}z_{k}^{2}\leq\eta,\!\forall k\in\mathcal{K}\label{eq:eesource-2}
\end{gather}
where $\{z_{k}\}_{k=1}^{K}$ are newly introduced variables. We remark
that function $z_{k}^{2}/\sqrt{q_{1k}}$ is convex (see Appendix \ref{sec:Convexity-of-Function}
for the proof), and so is (\ref{eq:eesource-2}). Also, (\ref{eq:eesource-1})
can be rewritten as $\frac{1}{r_{1k}}\leq\frac{z_{k}^{2}}{\tilde{\tau}}$
where the nonconvex part is quadratic-over-linear.

With the above transformations, (\ref{Prob:epiprob}) can be reformulated
as\begin{subequations}\label{Prob:epiprob-1}
\begin{align}
\underset{\substack{\mathbf{q},{\bf w},{\bf r},\tilde{\tau},\eta\\
\mathbf{v},\mathbf{s},\mathbf{u},\mathbf{t},\mathbf{z}
}
}{\text{minimize}}\  & \eta\label{eq:obj-1-1}\\
\text{subject to}\  & r_{1k}\ge(1+\tilde{\tau})Q_{1k},\,\forall k\in{\cal K}\label{eq:QoS-1}\\
 & q_{1k}\geq1/\bar{P}_{1k},\,\forall k\in{\cal K}\label{eq:source:txpower-1}\\
 & {\bf w}\herm{\bf A}_{l}{\bf w}+\sum_{k\in{\cal K}}\frac{{\bf w}\herm{\bf B}_{1kl}{\bf w}}{q_{1k}}\leq\bar{P}_{l}^{\text{R}},\,\forall l\in{\cal L}\label{eq:relaypow-1}\\
\; & \eqref{eq:rate-1},\eqref{eq:rate-2},\eqref{eq:rate-3},\eqref{eq:Econst-1},\eqref{eq:Econst-2},\eqref{eq:Econst-3},\eqref{eq:eesource-1},\eqref{eq:eesource-2}\label{eq:relaypower-1-1}
\end{align}
\end{subequations}where $\mathbf{q}\triangleq[q_{1k},...,q_{1K}]\trans$,
$\mathbf{v}\triangleq[v_{1},...,v_{K}]\trans$, $\mathbf{s}\triangleq[s_{1},...,s_{K}]\trans$,
$\mathbf{u}\triangleq[u_{1},...,u_{L}]\trans$, $\mathbf{t}\triangleq[t_{1},...,t_{L}]\trans$,
and $\mathbf{z}\triangleq[z_{1},...,z_{K}]\trans$; (\ref{eq:QoS-1}),
(\ref{eq:source:txpower-1}), and (\ref{eq:relaypow-1}) are respectively
the versions of (\ref{eq:QoS}), (\ref{eq:source:txpower}), and (\ref{eq:relaypow})
after change of variables. The equivalence here is in the sense of
optimality (see the proof in Appendix \ref{sec:Problem-Equivalences}).

We are now ready to use IA for solving (\ref{Prob:epiprob-1}). Specifically,
by applying the approximate formulations provided in Section \ref{subsec:AppForm}
to the nonconvex parts in (\ref{Prob:epiprob-1}), we obtain the following
convex approximation of (\ref{Prob:epiprob-1}) solved at iteration
$n+1$ \begin{subequations}\label{Prob:epiprob-1-1}
\begin{align}
\underset{\boldsymbol{\psi}}{\text{minimize}}\  & \eta\label{eq:obj-1-1-1}\\
\text{subject to\ } & h^{\textrm{bi}}(s_{k},v_{k};\frac{v_{k}^{(n)}}{s_{k}^{(n)}})+h^{\textrm{qol}}({\bf w},q_{1k};{\bf w}^{(n)},q_{1k}^{(n)};{\bf H}_{kk})\leq0,\,\forall k\in{\cal K}\label{eq:appr-ow-1}\\
 & \sum_{k\in{\cal K}}\frac{{\bf w}\herm{\bf B}_{1kl}{\bf w}}{q_{1k}}+h^{\textrm{po}}(u_{l};u_{l}^{(n)};2)\leq0,\,\forall l\in{\cal L}\label{eq:appr-ow-2}\\
 & \log(\tilde{\tau}-t_{l}-1)\geq h^{\textrm{log}}(\alpha_{l}t_{l};\alpha_{l}t_{l}^{(n)})+\sum_{k\in{\cal K}}c_{l}|f_{1kl}|^{2}h^{\textrm{po}}(q_{1k};q_{1k}^{(n)};-1),\!\forall l\in\mathcal{L}\label{eq:appr-ow-3}\\
 & \frac{1}{r_{1k}}+h^{\textrm{qol}}(z_{k},\tilde{\tau};z_{k}^{(n)},\tilde{\tau}^{(n)};1)\leq0,\!\forall k\in\mathcal{K}\label{eq:appr-ow-4}\\
\; & \eqref{eq:rate-1},\eqref{eq:rate-2},\eqref{eq:Econst-3},\eqref{eq:eesource-2},\eqref{eq:QoS-1},\eqref{eq:source:txpower-1},\eqref{eq:relaypow-1}\label{eq:relaypower-1-1-1}
\end{align}
\end{subequations} where $\bm{\psi}\triangleq[\mathbf{q}\trans,{\bf w}\trans,{\bf r}\trans,\tilde{\tau},\eta,\mathbf{v}\trans,\mathbf{s}\trans,\mathbf{u}\trans,\mathbf{t}\trans,\mathbf{z}\trans]\trans$
and $\boldsymbol{\psi}^{(n)}$ is some feasible point of (\ref{Prob:epiprob-1}).
\begin{algorithm}[h]
\caption{The Proposed Method Solving EEF-OW}
\label{Alg.SCA-ow}

\begin{algorithmic}[1]

\STATE \textbf{Initialization:} Set $n\coloneqq0$, $n'\coloneqq0$,
and randomly generate a feasible point $\boldsymbol{\psi}^{(0)}$
of (\ref{eq:obj-1-1-2}).

\REPEAT[Finding a feasible point of \eqref{Prob:epiprob-1}]

\STATE{Solve $\underset{\boldsymbol{\psi}\in\mathcal{S}(\boldsymbol{\psi}^{(n')})}{\text{minimize}}\ \eta+b\sum_{k\in{\cal K}}[(1+\tilde{\tau})Q_{1k}-r_{1k}]^{+}$,
denote the optimal by $\bm{\psi}_{\text{fe}}^{\ast}$.}\label{solveinitial}

\STATE{Update $n':=n'+1$, $\boldsymbol{\psi}^{(n')}:=\bm{\psi}_{\text{fe}}^{\ast}$.}

\UNTIL{$\sum_{k\in{\cal K}}[(1+\tilde{\tau}^{\ast})Q_{1k}-r_{1k}^{\ast}]^{+}=0$.}

\STATE Set $\boldsymbol{\psi}^{(0)}\coloneqq\boldsymbol{\psi}^{(n')}$.

\REPEAT[Solving \eqref{Prob:epiprob-1}]

\STATE{Obtain the optimal point of (\ref{Prob:epiprob-1-1}), denoted
by $\boldsymbol{\psi}^{\ast}$.}

\STATE{Update $n:=n+1$, $\boldsymbol{\psi}^{(n)}:=\boldsymbol{\psi}^{\ast}$.}

\UNTIL{convergence or predefined number of iterations.}

\STATE \textbf{Output} (solution for EEF-OW)\textbf{:} $\tau\coloneqq\frac{\tilde{\tau}^{(n)}-1}{\tilde{\tau}^{(n)}+1}$,
$\mathbf{w}\coloneqq\mathbf{w}^{(n)}$, $p_{1k}\coloneqq1/q_{1k}^{(n)}$
for all $k\in\mathcal{K}$.

\end{algorithmic}
\end{algorithm}

\subsubsection{Finding Initial Feasible Points\label{subsec:Finding-Initial-Feasible}}

A feasible point of (\ref{Prob:epiprob-1}) is required for starting
the IA procedure, which is difficult to find due to the QoS constraints.
Here we provide an efficient heuristic method inspired by \cite{Lipp:CCP:2016},\cite[Section 3.2]{advanceADC}
to overcome this issue. The idea is to allow the QoS constraints to
be violated, and the violation is penalized. Particularly, let us
consider the following modification of (\ref{Prob:epiprob-1})
\begin{align}
\underset{\boldsymbol{\psi}\in\mathcal{S}}{\text{minimize}}\  & \eta+b\sum_{k\in{\cal K}}[(1+\tilde{\tau})Q_{1k}-r_{1k}]^{+}\label{eq:obj-1-1-2}
\end{align}
where $b>0$ is a penalty parameter; $\mathcal{S}\triangleq\{\boldsymbol{\psi}|\eqref{eq:rate-1}\textrm{--}\eqref{eq:eesource-2},\eqref{eq:source:txpower-1},\eqref{eq:relaypow-1}\}$.
Finding feasible points of (\ref{eq:obj-1-1-2}) is easy as follows.
We first randomly generate $\tau^{(0)}\in(0,1)$, $0<p_{1k}^{(0)}\leq\bar{P}_{1k}$,
and ${\bf w}^{(0)}\in\mathbb{C}^{L\times1}$, then (if necessary)
scale ${\bf w}^{(0)}$ so that (\ref{eq:EnergyCons}) and (\ref{eq:relaypow})
are satisfied. Based on $(\tau^{(0)},p_{1k}^{(0)},{\bf w}^{(0)})$,
${\bf r}^{(0)},\mathbf{v}^{(0)},\mathbf{s}^{(0)},\mathbf{u}^{(0)},\mathbf{t}^{(0)}$,
and $\mathbf{z}^{(0)}$ are determined by setting (\ref{eq:rate-1}),
(\ref{eq:rate-2}), (\ref{eq:rate-3}), (\ref{eq:Econst-1}), (\ref{eq:Econst-3}),
and (\ref{eq:eesource-1}) to be equality. With $\boldsymbol{\psi}^{(0)}$,
we can start an iterative IA procedure for solving (\ref{eq:obj-1-1-2}).
Intuitively, the penalty term in (\ref{eq:obj-1-1-2}) would force
$\{(1+\tilde{\tau})Q_{k}-r_{1k}\}$ to decrease. Once $(1+\tilde{\tau})Q_{k}-r_{1k}\leq0$
for all $k$, i.e., the penalty term is zero, producing a feasible
point of (\ref{Prob:epiprob-1}).

In summary, we outline the proposed method for solving EEF-OW in Algorithm
\ref{Alg.SCA-ow}. In line \ref{solveinitial}, $\mathcal{S}(\boldsymbol{\psi}^{(n)})\triangleq\{\boldsymbol{\psi}|\eqref{eq:rate-1},\eqref{eq:rate-2},\eqref{eq:Econst-3},\eqref{eq:eesource-2},\eqref{eq:source:txpower-1},\eqref{eq:relaypow-1},\eqref{eq:appr-ow-1}\textrm{--}\eqref{eq:appr-ow-4}\}$
is an approximate convex set of $\mathcal{S}$ corresponding to $\boldsymbol{\psi}^{(n)}$.

\subsection{Solution for EEF-TW\label{subsec:Solution-for-EEF-TW}}

The procedure for finding a solution of EEF-TW is similar to the one
presented in the previous subsection. So, for the sake of brevity,
only the main steps are presented. We first arrive at the epigraph
form of EEF-TW given by\begin{subequations}\label{Prob:epiprob-2}\vspace{-7mm}

\begin{align}
\underset{\tilde{\mathbf{p}},{\bf w},\tilde{\mathbf{r}},\tau,\tilde{\eta}}{\text{minimize}}\  & \tilde{\eta}\label{eq:obj-1-2}\\
\text{subject to}\  & f_{k}^{\text{EE,TW}}(\tau,\tilde{\mathbf{p}},\tilde{\mathbf{r}})\geq\tilde{\eta}^{-1},\,\forall k\in{\cal K}\label{eq:eeconst-tw}\\
\; & \eqref{eq:rate-tw},\eqref{eq:EnergyCons-tw},\eqref{eq:QoS-tw},\eqref{eq:source:txpower-tw},\eqref{eq:relaypow-tw}.\label{eq:constraints-1}
\end{align}
\end{subequations}We focus on the nonconvex convexity induced by
(\ref{eq:rate-tw}), (\ref{eq:EnergyCons-tw}), (\ref{eq:relaypow-tw}),
and (\ref{eq:eeconst-tw}). Again, by using the change of variables
in Section \ref{subsec:Changes-of-Variables} and introducing additional
variables, we transform (\ref{Prob:epiprob-2}) into the following
equivalent problem \begin{subequations}\label{Prob:equivalent-tw}\vspace{-5mm}

\begin{align}
\underset{\substack{\tilde{\mathbf{q}},{\bf w},\tilde{\mathbf{r}},\tilde{\tau},\tilde{\eta}\\
\tilde{\mathbf{v}},\tilde{\mathbf{s}},\mathbf{u},\mathbf{t},\mathbf{z}
}
}{\text{minimize}}\  & \tilde{\eta}\label{eq:obj-1-2-1}\\
\text{subject to}\  & r_{ik}\ge(1+\tilde{\tau})Q_{ik},\,\forall k\in{\cal K},i=\{1,2\}\label{eq:qos-2-tw}\\
 & q_{ik}\geq1/\bar{P}_{ik},\,\forall k\in{\cal K},i=\{1,2\}\label{eq:sourcepow-2-tw}\\
 & {\bf w}\herm{\bf A}_{l}{\bf w}+\sum_{k\in{\cal K}}\sum_{i=1}^{2}\frac{{\bf w}\herm{\bf B}_{ikl}{\bf w}}{q_{ik}}\leq\bar{P}_{l}^{\text{R}},\,\forall l\in{\cal L}\label{eq:relaypow-2-tw}\\
 & r_{ik}\leq\log(1+v_{\bar{i}k}),\!\forall k\in{\cal K},i=\{1,2\},\bar{i}=\{1,2\}\setminus\{i\}\label{eq:rate-2-1-tw}\\
 & \sum_{j\in{\cal K}\backslash\{k\}}\sum_{\hat{i}=1}^{2}\frac{{\bf w}\herm\tilde{{\bf H}}_{ik\hat{i}j}{\bf w}}{q_{\hat{i}j}}+{\bf w}\herm{\bf G}_{ik}{\bf w}+\sigma^{2}\leq s_{ik},\!\forall k\in{\cal K},i=1,2\label{eq:rate-2-2-tw}\\
 & s_{ik}v_{ik}\leq\frac{{\bf w}\herm{\bf H}_{kk}{\bf w}}{q_{\bar{i}k}},\!\forall k\in{\cal K},i=\{1,2\},\bar{i}=\{1,2\}\setminus\{i\}\label{eq:rate-2-3-tw}\\
 & \sum_{k\in{\cal K}}\sum_{i=1}^{2}\frac{{\bf w}\herm{\bf B}_{ikl}{\bf w}}{q_{ik}}\leq u_{l}^{2},\!\forall l\in\mathcal{L}\label{eq:eh-2-tw}\\
 & \log(\tilde{\tau}-t_{l}-1)-\log(\alpha_{l}t_{l})+c_{l}\sum_{k\in{\cal K}}\sum_{i=1}^{2}\frac{|f_{ikl}|^{2}}{q_{ik}}\geq0,\!\forall l\in\mathcal{L}\label{eq:eh-2-2-tw}\\
 & \tilde{\varepsilon}_{l}\sqrt{{\bf w}\herm{\bf A}_{l}{\bf w}+u_{l}^{2}}\leq\hat{\beta}_{l}t_{l}-\bar{\beta}_{l}\tilde{\tau}+\check{\beta}_{l},\!\forall l\in\mathcal{L}\label{eq:eh-2-3-tw}\\
 & \frac{\tilde{\tau}}{\sum_{i=1}^{2}r_{ik}}\leq z_{k}^{2},\!\forall k\in{\cal K}\label{eq:ee-2-1-tw}\\
 & \left(\frac{1}{\sum_{i=1}^{2}r_{ik}}+z_{k}^{2}\right)E_{k}^{\textrm{TW,cir}}+\left(\sum_{i=1}^{2}\varepsilon_{ik}\frac{z_{k}^{2}}{\sqrt{q_{ik}}}\right)+\frac{\sum_{i=1}^{2}\rho_{ik}^{\textrm{sp}}r_{ik}}{\sum_{i=1}^{2}r_{ik}}\leq\tilde{\eta},\!\forall k\in{\cal K}\label{eq:ee-2-2-tw}
\end{align}
\end{subequations}where $\tilde{\mathbf{v}}\triangleq[v_{11},v_{21},...,v_{1K},v_{2K}]\trans$,
$\tilde{\mathbf{s}}\triangleq[s_{11},s_{21},...,s_{1K},s_{2K}]\trans$.
Similar to EEF-OW, the equivalence here is in the sense of optimality.
In (\ref{Prob:equivalent-tw}), the nonconvex parts include (\ref{eq:rate-2-3-tw}),
(\ref{eq:eh-2-tw}), (\ref{eq:eh-2-2-tw}), (\ref{eq:ee-2-1-tw}),
(\ref{eq:ee-2-2-tw}), which can also be approximated using the approximate
functions provided in Section \ref{subsec:AppForm}. By doing so,
we arrive at the convex approximation problem given as \begin{subequations}\label{Prob:equivalent-tw-1}\vspace{-3mm}
\begin{align}
 & \underset{\tilde{\boldsymbol{\psi}}}{\text{minimize}}\quad\tilde{\eta}\label{eq:obj-1-2-1-1}\\
 & \text{subject to}\ \nonumber \\
 & h^{\textrm{bi}}\left(s_{ik},v_{ik};v_{ik}^{(n)}/s_{ik}^{(n)}\right)+h^{\textrm{qol}}({\bf w},q_{\bar{i}k};{\bf w}^{(n)},q_{\bar{i}k}^{(n)};{\bf H}_{kk})\leq0,\forall k\in{\cal K},i=\{1,2\},\bar{i}=\{1,2\}\setminus\{i\}\label{eq:rate-2-2-tw-1}\\
 & \sum_{k\in{\cal K}}\sum_{i=1}^{2}\frac{{\bf w}\herm{\bf B}_{ikl}{\bf w}}{q_{ik}}+h^{\textrm{po}}(u_{l};u_{l}^{(n)};2)\leq0,\forall l\in\mathcal{L}\label{eq:eh-2-tw-1}\\
 & \log(\tilde{\tau}-t_{l}-1)\geq h^{\textrm{log}}(\alpha_{l}t_{l};\alpha_{l}t_{l}^{(n)})+\sum_{k\in{\cal K}}\sum_{i=1}^{2}c_{l}|f_{ikl}|^{2}h^{\textrm{po}}(q_{ik};q_{ik}^{(n)};-1),\forall l\in\mathcal{L}\label{eq:eh-2-2-tw-1}\\
 & \frac{1}{\sum_{i=1}^{2}r_{ik}}+h^{\textrm{qol}}(z_{k},\tilde{\tau};z_{k}^{(n)},\tilde{\tau}^{(n)};1)\leq0,\forall k\in{\cal K}\label{eq:ee-2-1-tw-1}\\
 & \left(\frac{1}{\sum_{i=1}^{2}r_{ik}}+z_{k}^{2}\right)E_{k}^{\textrm{TW,cir}}+\left(\sum_{i=1}^{2}\varepsilon_{ik}\frac{z_{k}^{2}}{\sqrt{q_{ik}}}\right)\nonumber \\
 & \hspace{1cm}+h^{\textrm{frac}}\left(\sum_{i=1}^{2}\rho_{ik}^{\textrm{sp}}r_{ik},\sum_{i=1}^{2}r_{ik};\frac{1}{(\sum_{i=1}^{2}\rho_{ik}^{\textrm{sp}}r_{ik}^{(n)})(\sum_{i=1}^{2}r_{ik}^{(n)})}\right)\leq\tilde{\eta},\forall k\in{\cal K}\label{eq:ee-2-2-tw-1}\\
 & \eqref{eq:qos-2-tw},\eqref{eq:sourcepow-2-tw},\eqref{eq:relaypow-2-tw},\eqref{eq:rate-2-1-tw},\eqref{eq:rate-2-2-tw},\eqref{eq:eh-2-3-tw}
\end{align}
\end{subequations}where $\tilde{\boldsymbol{\psi}}\triangleq[\tilde{\mathbf{q}}\trans,{\bf w}\trans,\tilde{\mathbf{r}}\trans,\tilde{\tau},\tilde{\eta},\tilde{\mathbf{v}}\trans,\tilde{\mathbf{s}}\trans,\mathbf{u}\trans,\mathbf{t}\trans,\mathbf{z}]\trans$
and $\tilde{\boldsymbol{\psi}}^{(n)}$ is a feasible point of (\ref{Prob:equivalent-tw}).
Finally, for finding initial feasible points of (\ref{Prob:equivalent-tw}),
we use a similar technique as that in Section \ref{subsec:Finding-Initial-Feasible}.

The proposed procedure for solving EEF-TW is outlined in Algorithm
\ref{Alg.SCA-tw}. In line \ref{solveinitial-tw}, $\tilde{\mathcal{S}}(\tilde{\boldsymbol{\psi}}^{(n)})\triangleq\{\tilde{\boldsymbol{\psi}}|\eqref{eq:qos-2-tw},\eqref{eq:sourcepow-2-tw},\eqref{eq:relaypow-2-tw},\eqref{eq:rate-2-1-tw},\eqref{eq:rate-2-2-tw},\eqref{eq:eh-2-3-tw},\eqref{eq:rate-2-2-tw-1}\textrm{--}\eqref{eq:ee-2-2-tw-1}\}$
is an inner convex approximation of $\tilde{\mathcal{S}}$ at $\tilde{\boldsymbol{\psi}}^{(n)}$.\vspace{-5mm}
\begin{algorithm}[h]
\caption{The Proposed Method Solving EEF-TW}
\label{Alg.SCA-tw}

\begin{algorithmic}[1]

\STATE \textbf{Initialization:} Set $n\coloneqq0$, $n'\coloneqq0$,
and randomly generate a point $\tilde{\boldsymbol{\psi}}^{(0)}\in\tilde{\mathcal{S}}\triangleq\{\tilde{\boldsymbol{\psi}}|\eqref{eq:sourcepow-2-tw}\textrm{--}\eqref{eq:ee-2-2-tw}\}$.

\REPEAT[Finding a feasible point of \eqref{Prob:equivalent-tw}]

\STATE{Solve $\underset{\tilde{\boldsymbol{\psi}}\in\tilde{\mathcal{S}}(\tilde{\boldsymbol{\psi}}^{(n)})}{\text{minimize}}\ \tilde{\eta}+b\sum_{k\in{\cal K}}\sum_{i=1}^{2}[(1+\tilde{\tau})Q_{ik}-r_{ik}]^{+}$,
and denote the optimal by $\tilde{\bm{\psi}}_{\text{fe}}^{\ast}$.}\label{solveinitial-tw}

\STATE{Update $n':=n'+1$, $\tilde{\boldsymbol{\psi}}^{(n')}:=\tilde{\bm{\psi}}_{\text{fe}}^{\ast}$.}

\UNTIL{$\sum_{k\in{\cal K}}\sum_{i=1}^{2}[(1+\tilde{\tau}^{\ast})Q_{ik}-r_{ik}^{\ast}]^{+}=0$.}

\STATE Set $\tilde{\boldsymbol{\psi}}^{(0)}\coloneqq\tilde{\boldsymbol{\psi}}^{(n')}$.

\REPEAT[Solving \eqref{Prob:equivalent-tw}]

\STATE{Obtain the optimal point of (\ref{Prob:equivalent-tw-1}),
denoted by $\tilde{\boldsymbol{\psi}}^{\ast}$.}

\STATE{Update $n:=n+1$, $\tilde{\boldsymbol{\psi}}^{(n)}:=\tilde{\boldsymbol{\psi}}^{\ast}$.}

\UNTIL{convergence or predefined number of iterations.}

\STATE \textbf{Output} (solution for EEF-TW)\textbf{:} $\tau\coloneqq\frac{\tilde{\tau}^{(n)}-1}{\tilde{\tau}^{(n)}+1}$,
$\mathbf{w}\coloneqq\mathbf{w}^{(n)}$, $p_{ik}\coloneqq1/q_{ik}^{(n)}$
for all $k\in\mathcal{K}$, $i=\{1,2\}$.

\end{algorithmic}
\end{algorithm}

\subsection{Convergence of Algorithms 1 and 2}

The general convergence analysis of the IA framework has been provided
in \cite{Beck2010}. Thus, we only need to examine the conditions
posted there for justifying the convergence of Algorithms \ref{Alg.SCA-ow}
and \ref{Alg.SCA-tw}. First, we recall that the approximate functions
provided in Section \ref{subsec:AppForm} satisfy (\ref{eq:IAcondition}),
which corresponds to \cite[Property A]{Beck2010}. In addition, the
feasible set of (\ref{Prob:epiprob-1}) and (\ref{Prob:equivalent-tw})
are compact and nonempty. Thus it is guaranteed that the objective
sequences $\{\eta^{(n)}\}_{n=0}^{\infty}$ (Algorithm 1) and $\{\tilde{\eta}^{(n)}\}_{n=0}^{\infty}$
(Algorithm 2) are nonincreasing and converge \cite[Corollary 2.3]{Beck2010}.

However, since objectives in (\ref{Prob:epiprob-1}) and (\ref{Prob:equivalent-tw})
are not strongly convex, the iterates $\{\boldsymbol{\psi}^{(n)}\}_{n=0}^{\infty}$
and $\{\tilde{\boldsymbol{\psi}}^{(n)}\}_{n=0}^{\infty}$ might not
converge. This issue can be overcome by using proximal terms, i.e.,
replacing objective of (\ref{Prob:epiprob-1-1}) and (\ref{Prob:equivalent-tw-1})
by $\eta+a||\boldsymbol{\psi}-\boldsymbol{\psi}^{(n)}||_{2}^{2}$
and $\tilde{\eta}+a||\tilde{\boldsymbol{\psi}}-\tilde{\boldsymbol{\psi}}^{(n)}||_{2}^{2}$,
respectively, with an arbitrary regularization parameter $a>0$ \cite{Dinh:SCP:2011}.
By doing so, the objective sequences $\{\eta^{(n)}\}_{n=0}^{\infty}$
and $\{\tilde{\eta}^{(n)}\}_{n=0}^{\infty}$ are strictly decreasing
and $||\boldsymbol{\psi}^{(n)}-\boldsymbol{\psi}^{(n+1)}||_{2}\rightarrow0$
, $||\tilde{\boldsymbol{\psi}}^{(n)}-\tilde{\boldsymbol{\psi}}^{(n+1)}||_{2}\rightarrow0$
\cite[Proposition 3.2]{Beck2010}, which come from the following relations
\vspace{-3mm}
\[
\eta^{(n)}-\eta^{(n+1)}\geq a||\boldsymbol{\psi}^{(n+1)}-\boldsymbol{\psi}^{(n)}||_{2}^{2},\,\tilde{\eta}^{(n)}-\tilde{\eta}^{(n+1)}\geq a||\tilde{\boldsymbol{\psi}}^{(n+1)}-\tilde{\boldsymbol{\psi}}^{(n)}||_{2}^{2}.
\]

\subsection{Conic Formulations for Approximate Subproblems}

The approximate subproblems (\ref{Prob:epiprob-1-1}) and (\ref{Prob:equivalent-tw-1})
are cast as generic convex programs due to the logarithmic functions
involved. Theoretically, these problems can be efficiently solved
using a general purpose interior-point solver. However, from the practical
perspective, it is more numerically efficient if we can arrive at
a more standard convex program, e.g.,\ conic quadratic or semidefinite
program \cite{BenNemi:book:LectModConv}. We observe from (\ref{Prob:epiprob-1-1})
and (\ref{Prob:equivalent-tw-1}) that the objectives and constraints
are linear or SOC-representable, except the constraints containing
the logarithmic functions. Hence, we are motivated to develop SOC-presentable
approximations for these constraints. Towards the goal, we present
a concave lower bound of the logarithmic function given as\vspace{-2mm}
\begin{equation}
\log x\geq\log x'+2-\frac{2\sqrt{x'}}{\sqrt{x}}\label{eq:loginequality}
\end{equation}
which holds for all $x>0,x'>0$.  Inequality $\eqref{eq:loginequality}$
can be justified as follows. Let us define $g(x;x')\triangleq\log x-\log x'-2+\frac{2\sqrt{x'}}{\sqrt{x}}$
for $x>0,x'>0$. We can easily prove that $g(x;x')\geq0$  by checking
the first-order derivative of $g(x;x')$ with respect to $x$, i.e.,
\[
\frac{\partial g(x;x')}{\partial x}=\frac{1}{x}-\frac{\sqrt{x}'}{x\sqrt{x}}=\frac{1}{x}\left(1-\frac{\sqrt{x}'}{\sqrt{x}}\right),
\]
which clearly indicates that $\frac{\partial g(x;x')}{\partial x}\geq0$
if $x\geq x'$, and $\frac{\partial g(x;x')}{\partial x}\leq0$ if
$x\leq x'$. Accordingly, $g(x;x')$ achieves the minimum at $x=x'$
with $g(x=x';x')=0$, and thus $g(x;x')\geq0$ for all $x>0,x'>0$
which validates (\ref{eq:loginequality}). Since (\ref{eq:loginequality})
is verified to fulfill the conditions in (\ref{eq:IAcondition}),
we can replace the constraint $\log x\geq y$ by\vspace{-2mm}
\begin{equation}
\log x'+2-\frac{2\sqrt{x'}}{\sqrt{x}}\geq y.\label{eq:applog}
\end{equation}
In the IA-based iterative procedure, $x'$ is the value of $x$ obtained
in the preceding iteration. We note that (\ref{eq:applog}) admits
the SOC-representation, i.e.,
\begin{equation}
{\textstyle \eqref{eq:applog}}\Leftrightarrow\begin{cases}
\xi^{2}\leq x\\
\left\Vert [2\sqrt[4]{x'},\,\log x'+2-y,\,\xi]\right\Vert _{2}\leq\log x'+2-y+\xi
\end{cases}.
\end{equation}
In the same way, (\ref{eq:rate-1}) can be approximated by\vspace{-3mm}

\begin{equation}
\log(1+v_{k}^{(n)})+2-\frac{2\sqrt{1+v_{k}^{(n)}}}{\sqrt{1+v_{k}}}\ge r_{k}\,\forall k\in{\cal K}.
\end{equation}

\section{Designs Based on Zero-Forcing Beamforming\label{sec:Computational-Complexity}}

In multi-pair relay systems, ZF is commonly invoked to eliminate the
inter-pair interference, and thus, reduces the design complexity \cite{Tao2012TSP,Rankov2007JSAC,Joung2010TSP}.
For EEF-OW and EEF-TW, using ZF beamforming does not lead to convex
formulations due to the complexity involved. However, we can obtain
suboptimal solutions but with much lowered complexity, using the similar
procedures illustrated in Section \ref{sec:Proposed-Algorithms}.
In the rest of the section, we sequentially present the ZF-based designs
for EEF-OW and EEF-TW.\vspace{-5mm}

\subsection{ZF-Based Design for EEF-OW}

Let us define $\bar{\mathbf{H}}_{k}\triangleq[\mathbf{h}_{k1}\trans,...,\mathbf{h}_{k(k-1)}\trans,\mathbf{h}_{k(k+1)}\trans,...,\mathbf{h}_{kK}\trans]\in\mathbb{C}^{L\times(K-1)}$
and $\bar{\mathbf{H}}\triangleq[\bar{\mathbf{H}}_{1},...,\bar{\mathbf{H}}_{K}]\trans\in\mathbb{C}^{L\times K(K-1)}$.
The ZF beamforming principles lead to \vspace{-1mm}
\begin{equation}
\mathbf{h}_{kj}{\bf w}=0,\forall j\neq k,k\in\mathcal{K}\Leftrightarrow\bar{\mathbf{H}}{\bf w}=\mathbf{0}.
\end{equation}
Clearly, the null-space of $\bar{\mathbf{H}}$ exists if $L>K(K-1)$.
Let $\mathbf{Z}\in\mathbb{C}^{L\times(L-K(K-1))}$ be an orthogonal
basis of the null-space of $\bar{\mathbf{H}}$, then we can find ${\bf w}$
such as ${\bf w}=\mathbf{Z}\bar{{\bf w}}$ where $\bar{{\bf w}}\in\mathbb{C}^{(L-K(K-1))\times1}$
\cite{ZF_BD}. This allows us to rewrite SINR at $\mathrm{U}_{2k}$
as
\begin{align}
\gamma_{2k}^{\textrm{OW,ZF}}(\bar{{\bf w}},{\bf p}) & =\frac{p_{1k}\bar{{\bf w}}\herm\mathbf{H}_{kk}^{\textrm{ZF}}\bar{{\bf w}}}{\bar{{\bf w}}\herm\mathbf{G}_{2k}^{\textrm{ZF}}\bar{{\bf w}}+\sigma^{2}}
\end{align}
where $\mathbf{H}_{kk}^{\textrm{ZF}}\triangleq\mathbf{Z}\herm\mathbf{H}_{kk}\mathbf{Z}$
and $\mathbf{G}_{2k}^{\textrm{ZF}}\triangleq\mathbf{Z}\herm\mathbf{G}_{k}\mathbf{Z}$.
Thus, the design problem with ZF beamforming is \begin{subequations}\label{Prob:nullsapce-ow}\vspace{-3mm}
\begin{align}
\underset{{\bf p},\bar{{\bf w}},{\bf r},\tau}{\text{maximize}}\  & \min_{1\leq k\leq K}\ f_{k}^{\text{EE,OW}}(\tau,{\bf p},{\bf r})\label{eq:obj-2}\\
\text{subject to}\  & \bar{{\bf w}}\herm{\bf A}_{l}^{\textrm{ZF}}\bar{{\bf w}}+\sum_{k\in{\cal K}}p_{1k}\bar{{\bf w}}\herm{\bf B}_{1kl}^{\textrm{ZF}}\bar{{\bf w}}\leq\bar{P}_{l}^{\text{R}},\,\forall l\in{\cal L}\label{eq:relaypow-2}\\
 & r_{1k}\leq\log(1+\gamma_{2k}^{\textrm{OW,ZF}}(\bar{{\bf w}},{\bf p})),\,\forall k\in{\cal K}\\
 & \frac{1-\tau}{2}\tilde{\varepsilon}_{l}\sqrt{\bar{{\bf w}}\herm{\bf A}_{l}^{\textrm{ZF}}\bar{{\bf w}}+\sum_{k\in{\cal K}}p_{1k}\bar{{\bf w}}\herm{\bf B}_{1kl}^{\textrm{ZF}}\bar{{\bf w}}}+E_{l}^{\textrm{R,const}}\leq E_{l}^{\textrm{EH,OW}}(\tau,{\bf p}),\,\forall l\in{\cal L}.\\
 & \eqref{eq:QoS},\eqref{eq:source:txpower}.
\end{align}
\end{subequations}where ${\bf A}_{l}^{\textrm{ZF}}\triangleq\mathbf{Z}\herm\mathbf{A}_{l}\mathbf{Z}$
and ${\bf B}_{1kl}^{\textrm{ZF}}\triangleq\mathbf{Z}\herm\mathbf{B}_{1kl}\mathbf{Z}$.\vspace{-3mm}

\subsection{ZF-Based Design for EEF-TW}

To obtain ZF beamforming for two-way system, we first recall that
$\tilde{{\bf h}}_{ik\hat{i}j}=\tilde{{\bf h}}_{\hat{i}jik}$ and define
$\mathbf{M}_{k}\triangleq[\tilde{{\bf h}}_{1k1(k+1)}\trans,\tilde{{\bf h}}_{2k2(k+1)}\trans,...,\tilde{{\bf h}}_{1k1K}\trans,\tilde{{\bf h}}_{2k2K}\trans]$
and $\bar{\mathbf{M}}\triangleq[\mathbf{M}_{1},...,\mathbf{M}_{K-1},\bar{\mathbf{H}}_{1},...,\bar{\mathbf{H}}_{K}]\trans\in\mathbb{C}^{2K(K-1)\times L}$.
Then we can write the ZF constraint as
\begin{equation}
\tilde{{\bf h}}_{ik\hat{i}j}{\bf w}=0,\forall j\neq k,k\in\mathcal{K},i,\hat{i}\in\{1,2\}\Leftrightarrow\bar{\mathbf{M}}{\bf w}=\mathbf{0}.
\end{equation}
Let $\tilde{\mathbf{Z}}\in\mathbb{C}^{L\times(L-2K(K-1))}$ be an
orthogonal basis of null-space of $\bar{\mathbf{M}}$, which requires
the condition that $L>2K(K-1)$ for existence. Again, we can find
beamforming vector as ${\bf w}=\tilde{\mathbf{Z}}\bar{\mathbf{w}}$
where $\bar{\mathbf{w}}\in\mathbb{C}^{(L-2K(K-1))\times1}$. The SINR
at $\mathrm{U}_{ik}$ reduces to
\begin{align}
\gamma_{ik}^{\textrm{TW,ZF}}(\bar{{\bf w}},\tilde{\mathbf{p}}) & =\frac{p_{\bar{i}k}\bar{{\bf w}}\herm\tilde{\mathbf{H}}_{kk}^{\textrm{ZF}}\bar{{\bf w}}}{\bar{{\bf w}}\herm\tilde{\mathbf{G}}_{ik}^{\textrm{ZF}}\bar{{\bf w}}+\sigma^{2}}
\end{align}
where $\tilde{\mathbf{H}}_{kk}^{\textrm{ZF}}\triangleq\tilde{\mathbf{Z}}\herm\mathbf{H}_{kk}\tilde{\mathbf{Z}}$
and $\tilde{\mathbf{G}}_{ik}^{\textrm{ZF}}\triangleq\tilde{\mathbf{Z}}\herm{\bf G}_{ik}\tilde{\mathbf{Z}}$.
The EEF design problem based on ZF beamforming is given by \begin{subequations}\label{Prob:general-tw-zf}
\begin{align}
\underset{\tilde{\mathbf{p}},\bar{{\bf w}},\tilde{\mathbf{r}},\tau}{\text{maximize}}\  & \min_{1\leq k\leq K}\ f_{k}^{\text{EE,TW}}(\tau,\tilde{\mathbf{p}},\tilde{\mathbf{r}})\label{eq:obj-tw-1}\\
\text{subject to}\  & \bar{{\bf w}}\herm\tilde{{\bf A}}_{l}^{\textrm{ZF}}\bar{{\bf w}}+\sum_{k\in{\cal K}}\sum_{i=1}^{2}p_{ik}\bar{{\bf w}}\herm\tilde{{\bf B}}_{ikl}^{\textrm{ZF}}\bar{{\bf w}}\leq\bar{P}_{l}^{\text{R}},\,\forall l\in{\cal L}\label{eq:relaypow-tw-1}\\
 & r_{ik}\leq\log(1+\gamma_{ik}^{\textrm{TW,ZF}}(\bar{{\bf w}},\tilde{\mathbf{p}})),\forall k\in{\cal K},i=\{1,2\}.\\
 & \frac{1-\tau}{2}\tilde{\varepsilon}_{l}\sqrt{\bar{{\bf w}}\herm\tilde{{\bf A}}_{l}^{\textrm{ZF}}\bar{{\bf w}}+\sum_{k\in{\cal K}}\sum_{i=1}^{2}p_{ik}\bar{{\bf w}}\herm\tilde{{\bf B}}_{ikl}^{\textrm{ZF}}\bar{{\bf w}}}+E_{l}^{\textrm{R,const}}\leq E_{l}^{\textrm{EH,TW}}(\tau,\tilde{\mathbf{p}}),\,\forall l\in{\cal L}\\
 & \eqref{eq:QoS-tw},\eqref{eq:source:txpower-tw}.
\end{align}
\end{subequations}where $\tilde{{\bf A}}_{l}^{\textrm{ZF}}\triangleq\tilde{\mathbf{Z}}\herm\mathbf{A}_{l}\tilde{\mathbf{Z}}$
and $\tilde{{\bf B}}_{ikl}^{\textrm{ZF}}\triangleq\tilde{\mathbf{Z}}\herm\mathbf{B}_{ikl}\tilde{\mathbf{Z}}$.
\begin{rem}
We note that in ZF-based designs, other parameters (transmit data
rate, users\textquoteright{} transmit power, and EH time) are still
jointly designed with the ZF beamforming. Here, problems (\ref{Prob:nullsapce-ow})
and (\ref{Prob:general-tw-zf}) can be solved by the similar IA procedures
described in Sections \ref{sec:Proposed-Algorithms}. The two problems
are optimized over $\bar{{\bf w}}$. Thus, the total numbers of variables
in their convex approximate programs are smaller than those of EEF-OW
and EEF-TW (as discussed in the next section). On the other hand,
since the inter-pair interference is canceled, it is expected that
the numbers of iterations of IA procedures solving (\ref{Prob:nullsapce-ow})
and (\ref{Prob:general-tw-zf}) are smaller compared to those of EEF-OW
and EEF-TW. This will be elaborated by numerical experiments provided
in Subsection \ref{subsec:Performance-of-ZF}. For the ease of exposition,
we refer to the solutions of (\ref{Prob:nullsapce-ow}) and (\ref{Prob:general-tw-zf})
as `ZF-based design (OW)' and `ZF-based design (TW)', respectively.\vspace{-3mm}
\end{rem}

\section{Computational Complexity Analysis\label{sec:Computational-Complexity-Analysi}}

We now discuss on the computational complexity of solving the SOCP
approximations (in each of iterations) by a general interior point
method based on the results in \cite[Chapter 6]{BenNemi:book:LectModConv}.

For Algorithm 1, the SOCP solved at an iteration includes $(10K+6L+2)$
real variables and $(10K+5L)$ conic constraints. Then the worst case
of computational complexity in an iteration of the algorithm is $\mathcal{O}\bigl((10K+5L)^{0.5}(10K+6L)^{3}\bigr)$.
For Algorithm 2, the SOCP solved at an iteration includes $(19K+6L+2)$
real variables and $(16K+5L)$ conic constraints. Then the worst case
of computational complexity in an iteration of the algorithm is $\mathcal{O}\bigl((16K+5L)^{0.5}(19K+6L)^{3}\bigr)$,
which is higher than that of Algorithm 1 due to the additional variables
coming from the bi-directional transmission.

For ZF-based design (OW), by using ZF beamforming at the relays, the
number of real variables in an SOCP approximation is $(10K+5L+2-2K^{2})$
and the number of conic constraints is $(9K+5L)$. Hence the worst
case complexity estimate is $\mathcal{O}\bigl((9K+5L)^{0.5}(10K+5L-2K^{2})^{3}\bigr)$.
Similarly, for ZF-based design (TW), an SOCP approximation includes
$(19K+5L+2-2K^{2})$ real variables and $(14K+5L)$ conic constraints.
So, the complexity is $\mathcal{O}\bigl((14K+5L)^{0.5}(19K+5L-2K^{2})^{3}\bigr)$.

From the above complexity estimates, it is expected that computational
complexity in an iteration of ZF-based design (OW) and ZF-based design
(TW) are lower than that of EEF-OW and EEF-TW, respectively. This
point will be numerically elaborated in Table \ref{Tab2}. \vspace{-3mm}

\section{Numerical Results\label{sec:Simulation-Result}}

In this section, we numerically evaluate the proposed methods. We
consider a relay network as depicted in Fig. 1 in which the distance
between two users of each pair is 10 m. The relays are randomly placed
inside the rectangular region formed by the users $\{\mathrm{U}_{1k}\}_{k=1}^{K}$
and $\{\mathrm{U}_{2k}\}_{k=1}^{K}$. The exponent path loss model
is used with path loss exponent 3.5. All channels are Rayleigh fading.
Simulation parameters are taken from Table \ref{Tab. 1}, unless stated
otherwise. The maximum transmit power is set to be the same for all
users, i.e., $\bar{P}_{ik}=\bar{P}$, $\forall i,k$, which varies
in the experiments. The number of user pairs is $K=3$. Other parameters
will be specified in the experiments. In all simulations, the iterative
procedures of Algorithms 1 and 2 stop when either the increase in
the objective between two consecutive iterations is less than $10^{-5}$
or the number of iterations exceeds 200. To solve convex problems,
we use the MOSEK \cite{Mosek} and Fmincon solvers in MATLAB environment.
\begin{table}[tb]
\caption{Simulation Parameters}

\centering{}%
\begin{tabular}{c|c||c|c}
\hline
\noun{Parameters} & \noun{Value} & \noun{Parameters} & \noun{Value}\tabularnewline
\hline
\hline
Bandwidth  & 250 kHz & User circuit power & $P_{ik}^{\text{idle}}=0.1$mW, $P_{ik}^{\text{act,cir}}=1$ mW\tabularnewline
\hline
Noise power & $\sigma^{2}=\tilde{\sigma}^{2}=-90$ dBm & Relay circuit power & $P_{l}^{\text{R,const}}=1$ mW\tabularnewline
\hline
QoS & $Q_{ik}=0.5$ nats/s/Hz & Signal processing power \cite{Cui2017WCOM} & $\rho_{ik}^{\text{en}}=\rho_{ik}^{\text{de}}=50$ mW/(Gnats/s)\tabularnewline
\hline
\multirow{2}{*}{PA model \cite{Cui2017WCOM}} & $\bar{P}_{l}^{\text{R}}=33$ dBm,  & \multirow{2}{*}{EH model \cite{Boshkovska2015COML}} & $\bar{P}_{l}^{\text{DC}}=24$ mW,\tabularnewline
 & $\epsilon_{ik}=\epsilon_{l}^{R}=0.35$ &  & $c_{l}=150$, $d_{l}=0.014$\tabularnewline
\hline
\end{tabular}\label{Tab. 1}\vspace{-5mm}
\end{table}
\begin{figure}
\centering{}\subfigure[Convergence over two channel realizations]{\label{Fig.2.a}\vspace{1mm}\includegraphics[width=0.46\columnwidth]{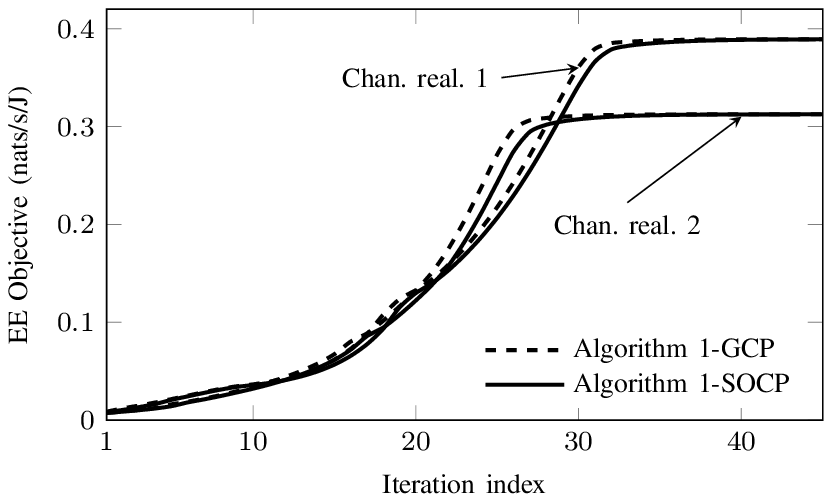}}\quad{}\subfigure[CDFs of the number of required iterations to converge.]{\label{Fig.2.b}\includegraphics[width=0.47\columnwidth]{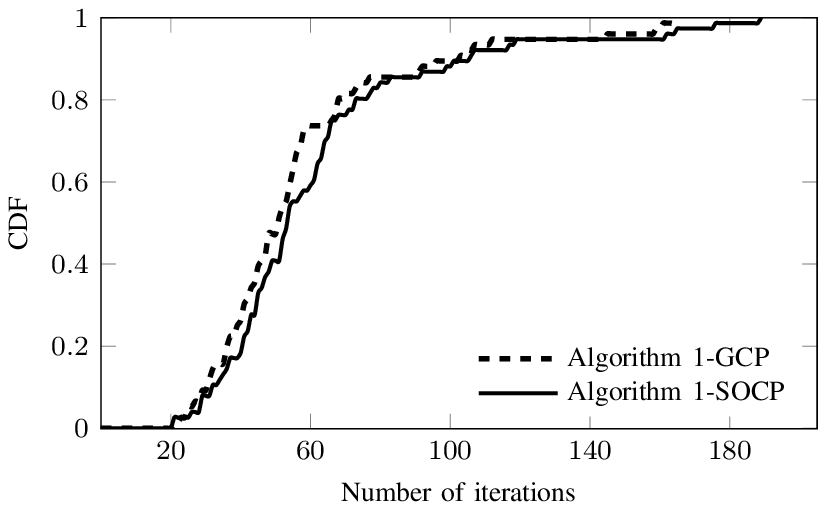}}\vspace{-3mm}\caption{Impact of conic formulation on the convergence behavior of Algorithm
\ref{Alg.SCA-ow}. We take $K=2$, $L=2$ and $\bar{P}=33$ dBm.}
\label{Fig.2}\vspace{-6mm}
\end{figure}
\vspace{-4mm}

\subsection{Performances of Algorithm \ref{Alg.SCA-ow} (One-Way Relaying)\label{subsec:Sim:OWR}}

In the first set of experiments, we study the impact of conic formulation
of (\ref{Prob:epiprob-1-1}) on the computational complexity of Algorithm
1. Fig. \ref{Fig.2} shows the convergence behavior of Algorithm~1
running with the generic convex program (GCP) and SOCP. Specifically,
Fig. \ref{Fig.2.a} plots the convergence of the objective over two
channel realizations, and Fig. \ref{Fig.2.b} shows the cumulative
distribution functions (CDFs) of the required number of iterations
to converge. Also, we provide the average total and per-iteration
run time of the algorithm with the two formulations in Table \ref{Tab:Runtime}.
We can see in the figure that with the SOCP, the algorithm converges
with more iterations compared to the GCP. However, as shown in the
table, the per-iteration run time of the SOCP (solver Fmincon) is
much smaller than that of GCP (solver Fmincon), resulting that the
total run time of the algorithm with the SOCP is ten times smaller
than that with the GCP. In addition, the SOCP allows us to use the
more efficient solver MOSEK. With this, the total run time significantly
reduces.
\begin{table}
\centering{}\caption{Average per-iteration and total solver run time (in second) of Algorithm
1 adopting GCP and SOCP. We take $K=2$, $L=2$ and $\bar{P}=33$
dBm.}
\label{Tab:Runtime}\vspace{-3mm}%
\begin{tabular}{c|c||c|c}
\hline
\multicolumn{2}{c||}{Solver} & Fmincon & MOSEK\tabularnewline
\hline
\hline
\multirow{2}{*}{Algorithm1-GCP} & Avg. per-iteration run time & 49 & \multirow{2}{*}{N/A}\tabularnewline
 & Avg. total run time & 2.5e3 & \tabularnewline
\hline
\multirow{2}{*}{Algorithm1-SOCP} & Avg. per-iteration run time & 4.97 & 0.003\tabularnewline
 & Avg. total run time & 220 & 0.17\tabularnewline
\hline
\end{tabular}\vspace{-4mm}
\end{table}
\begin{figure}[t]
\centering{}\subfigure[Convergence over a channel realization]{\label{Fig.3.a}\includegraphics[width=0.47\columnwidth]{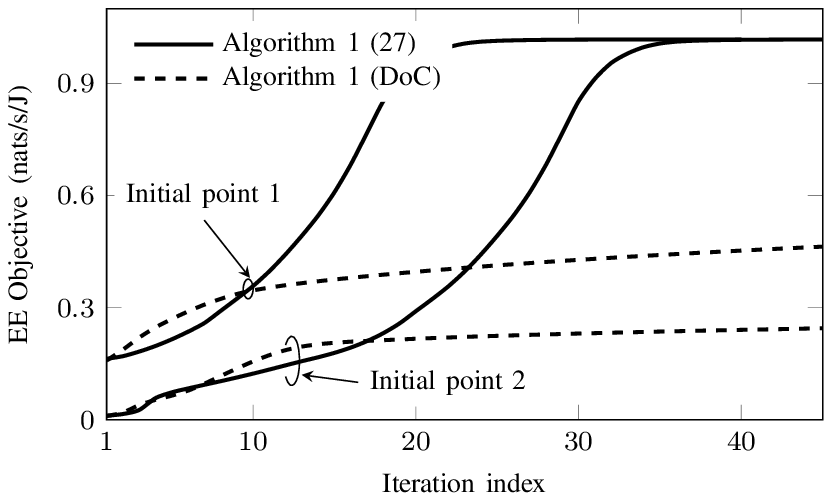}}\quad{}\subfigure[CDF of the number of requirediterations to converge.]{\label{Fig.3.b}\includegraphics[width=0.47\columnwidth]{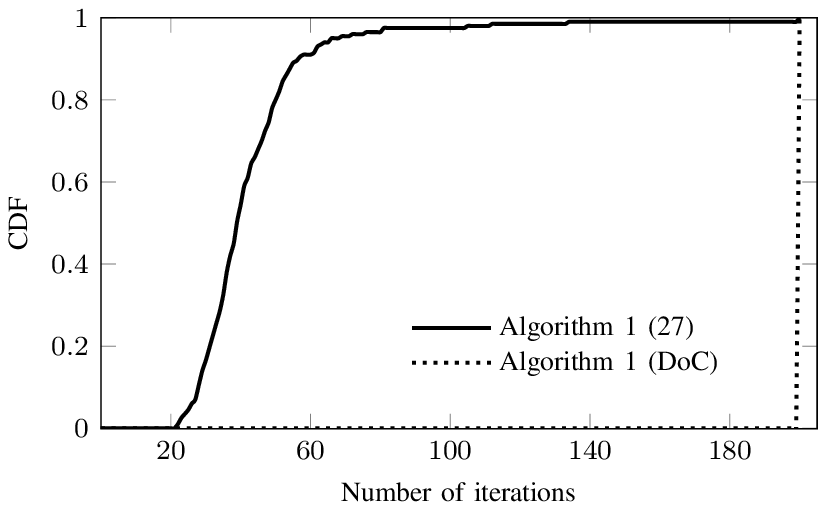}}\vspace{-2mm}\caption{Impact of approximation functions (\ref{eq:Biappr}) and DoC on the
convergence of Algorithm \ref{Alg.SCA-ow}. We take $K=3$, $L=9$
and $\bar{P}=33$ dBm.}
\label{Fig.3}\vspace{-3mm}
\end{figure}

In Fig. 3, we illustrate the effectiveness of (\ref{eq:Biappr}) in
term of convergence. Particularly, Fig. \ref{Fig.3.a} plots the convergence
of Algorithm \ref{Alg.SCA-ow} using the two approximation functions,
(\ref{eq:Biappr}) and difference-of-convex (DoC) function, over a
random channel realization with two different initial points generated
randomly. And Fig. \ref{Fig.3.b} shows the CDFs of number of iteration
required for convergence. The results clearly demonstrate that using
DoC formulations of bilinear function for the considered problems
is not efficient since the corresponding iterative procedure only
stops by the maximum number of iteration criteria. This confirms the
use of (\ref{eq:Biappr}).

\begin{figure}
\centering{}\includegraphics[width=0.5\columnwidth]{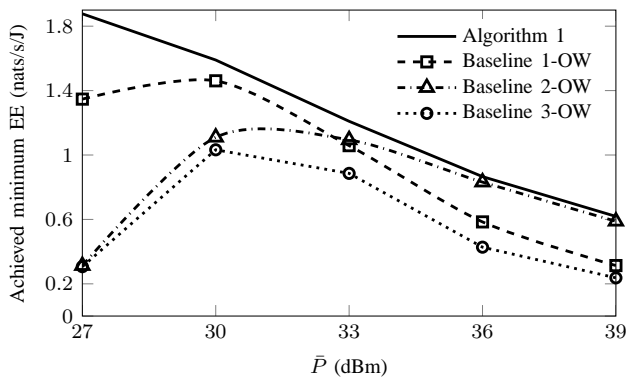}\vspace{-6mm}\caption{Achieved EE versus the transmit power $\bar{P}$ with $K=3$ and $L=9$.}
\label{Fig.3-1}\vspace{-6mm}
\end{figure}

Fig. \ref{Fig.3-1} depicts the averaged minimum EE performance of
Algorithm \ref{Alg.SCA-ow} as a function of the maximum user power
$\bar{P}$. For comparison purpose, we also provide the performance
of three baseline schemes: in the first scheme, namely `Baseline
1-OW', the transmit power of the users are fixed at $\bar{P}$; in
the second scheme, namely `Baseline 2-OW', EH time is fixed as $\tau=\frac{1}{3}$;
in the third scheme, namely `Baseline3 -OW', the users' transmit
power and EH time are fixed at $\bar{P}$ and $\tau=\frac{1}{3}$,
respectively. For the three baseline schemes, it may happen that feasible
resource allocation cannot be obtained for some channel realizations.
Thus, we set the performance of those infeasible channels as zero.
The first observation is that the performance of Algorithm 1 decreases
when $\bar{P}$ increases. This result can be explained as follows.
In an EE problem, the transmit power may be smaller than the threshold,
especially when the threshold is relatively large. For this case,
increasing $\bar{P}$ brings no benefit to the optimizing of the transmit
power. On the other hand, as shown in (\ref{eq:ampeff}), both $\bar{P}$
and the optimized transmit power influence the PA efficiency. And
increasing $\bar{P}$ reduces the PA efficiency, leading to more amount
of energy consumed at the PA as can be seen in (\ref{eq:realpowamp}).
Another interesting observation is that the EEs of the three baseline
schemes first increase, and then decrease as $\bar{P}$ increases.
The reason is that  the probabilities of infeasibility of these schemes
are high when $\bar{P}$ is small. When $\bar{P}$ becomes larger,
the infeasibility probabilities are smaller leading to the improved
performances. When the probabilities of infeasibility are small enough,
further increasing $\bar{P}$ leads to the degraded performances due
to the decrease of PA efficiency. As expected, our proposed scheme
outperforms the baseline ones.

\begin{figure}
\centering{}\begin{minipage}{0.46\columnwidth}\centering\includegraphics[width=0.98\columnwidth]{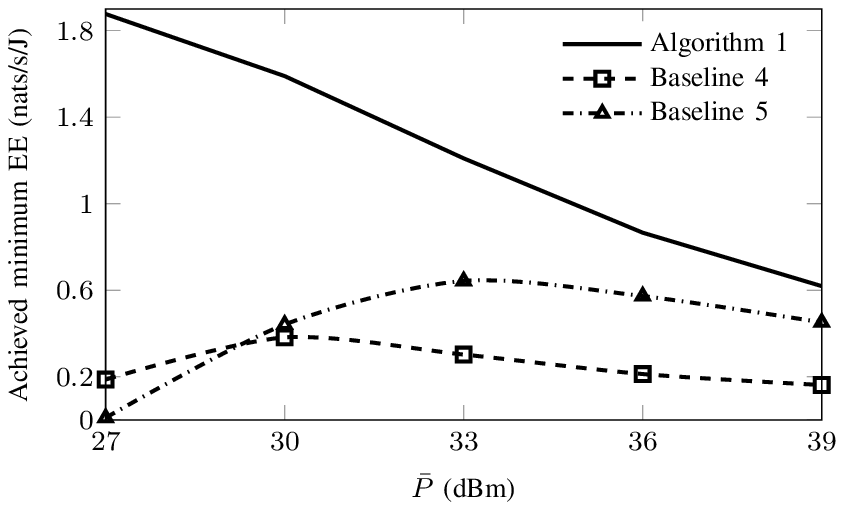}\vspace{-5mm}\caption{Achieved minimum EE and versus the transmit power $\bar{P}$ with
$K=3$ and $L=9$.}
\label{Fig.4}\end{minipage}\qquad{}\begin{minipage}{0.46\columnwidth}\centering\includegraphics[width=0.98\columnwidth]{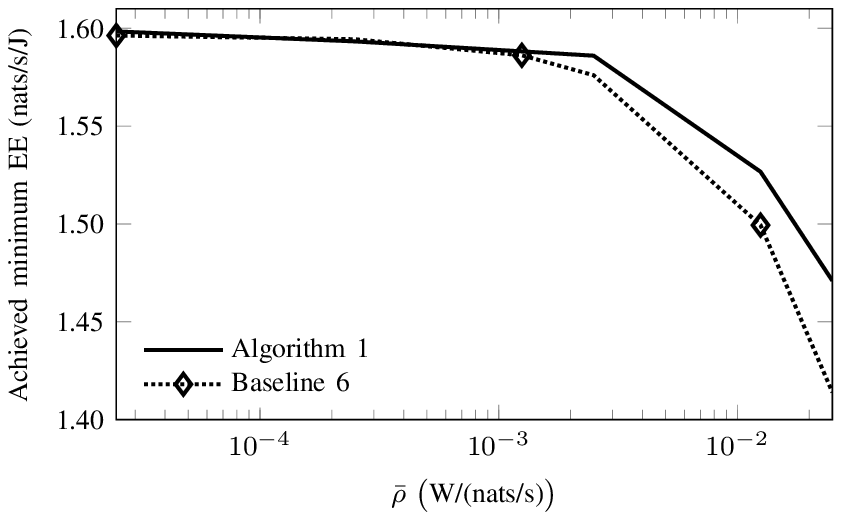}\vspace{-5mm}\caption{Achieved minimum EE versus rate-dependent power coefficient $\bar{\rho}$
with $K=3$ and $L=9$.}
\label{Fig.5}\end{minipage}\vspace{-8mm}
\end{figure}
In Fig. \ref{Fig.4}, we show the impacts of PA and EH models on the
minimum EE performance. For this purpose, we consider the following
schemes: the first scheme, named as `Baseline 4', considers linear
model of PA efficiency where the efficiency is fixed at 0.35. The
the second scheme, named as `Baseline 5', adopts the linear EH model
with constant conversion efficiency 0.8. The performances of these
schemes are obtained as follows. First, the design parameters are
determined by suitably modifying Algorithm 1 corresponding to the
considered models. From the achieved values, the minimum EE is recalculated
following the PA and EH models considered in Section \ref{sec:System-Model-and}.
If there is infeasibility, the corresponding minimum EE is set as
zero. The figure clearly shows that PA and EH models have significant
influence on the performance. Similarly to Baselines 1, 2 and 3 (in
Fig. \ref{Fig.3}), the performances of Baseline 4 and Baseline 5
are inferior when $\bar{P}$ is small due to high probability of infeasibility.
The performance degradations of Baselines 4 and 5 are mainly because
of the mismatch between the baseline schemes and the realistic models.
The results again confirm the validity of our proposed scheme.

To investigate the impacts of rate-dependent signal processing power
(RSPP) on the minimum EE performance, we let the rate-dependent-power
coefficients in each pair be different from that of other pairs by
simply setting as $\rho_{1k}^{\text{en}}=\rho_{1k}^{\text{de}}=\omega_{k}\bar{\rho}$
where $\omega_{k}=k$, and plot the performance as a function of $\bar{\rho}$
in Fig. \ref{Fig.5}. Here, the compared scheme, namely `Baseline
6', takes $\rho_{ik}^{\text{en}}=\rho_{ik}^{\text{de}}=0$, and its
performance is obtained similarly as that of Baseline 4 and Baseline
5 (in Fig. \ref{Fig.4}). We observe that RSPP has insignificant influence
on the performance when its coefficients are small. However, when
the coefficients becomes larger, the gap between Algorithm 1 and Baseline
6 is remarkable.
\begin{figure}
\centering{}\subfigure[Average individual EE of user pairs.]{\label{fig. fairness-OW-1}\includegraphics[width=0.47\columnwidth]{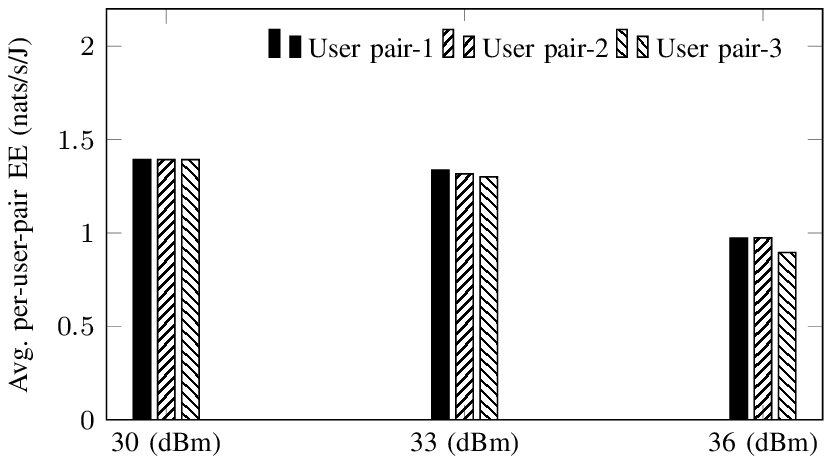}}\quad{}\subfigure[CDF of Jain's fairness index.]{\label{fig. fairness-OW-2}\includegraphics[width=0.49\columnwidth]{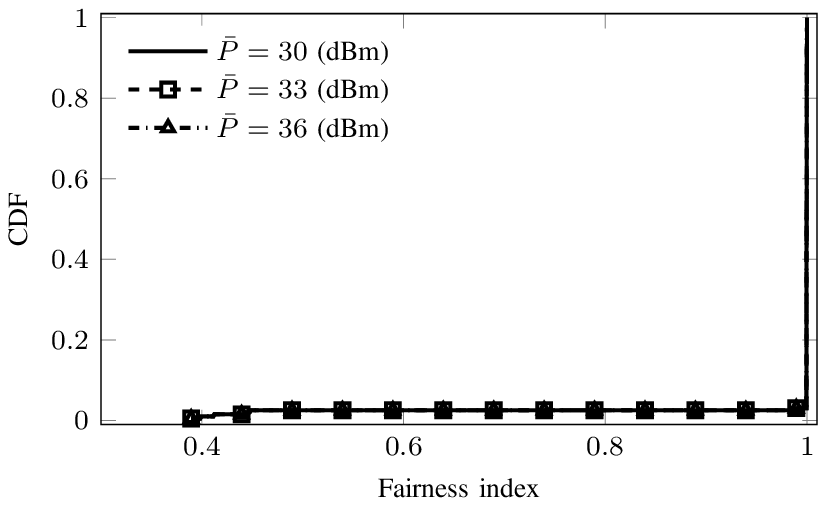}}\vspace{-3mm}\caption{EE fairness among the user pairs achieved by Algorithm 1 with $K=3$,
and $L=12$.}
\label{fig. fairness-OW}\vspace{-8mm}
\end{figure}

Fig. \ref{fig. fairness-OW} shows the EE fairness among user pairs
versus different values of $\bar{P}$. In particular, the average
individual EE of the user pairs is plotted in Fig. \ref{fig. fairness-OW-1},
and the CDFs of Jain's fairness index \cite{jain1984quantitative}\footnote{According to \cite{jain1984quantitative}, let us denote $[\text{EE}_{1}^{\ast},\ldots\text{EE}_{K}^{\ast}]$
as the individual EEs of the user pairs, then the fairness index is
given as: $\text{fairness}=\frac{\bigl(\sum_{k=1}^{K}\text{EE}_{k}^{\ast}\bigr)^{2}}{K\sum_{k=1}^{K}(\text{EE}_{k}^{\ast})^{2}}$.
Obviously, when $\text{EE}_{1}^{\ast}=\ldots=\text{EE}_{K}^{\ast}$,
$\text{fairness}=1$ which implies an absolute fairness. } are shown in Fig. \ref{fig. fairness-OW-2}. It can be observed that
the achieved EE is relatively balanced among all user pairs. On the
other hand, the algorithm achieves absolute fairness in more than
90\% of channel realizations in all considered cases of $\bar{P}$.\vspace{-3mm}

\subsection{Performances of Algorithm \ref{Alg.SCA-tw} (Two-Way Relaying)}

\begin{figure}
\centering
\subfigure[Convergence of Algorithm 2 for one channel realization with $\bar{P}=33$ dBm.]{\label{fig.6a}\includegraphics[width=0.46\columnwidth]{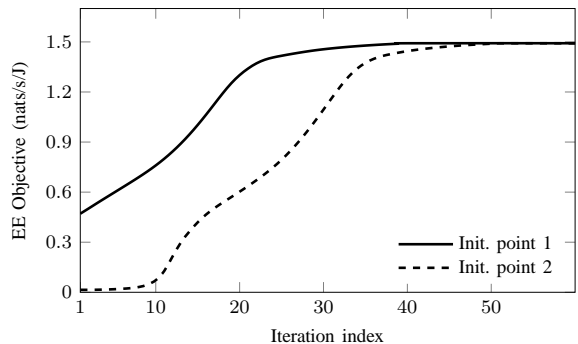}}\qquad{}
\subfigure[Achieved minimum EE performance versus the transmit power $\bar{P}$.]{ \label{fig.6b}\includegraphics[width=0.46\columnwidth]{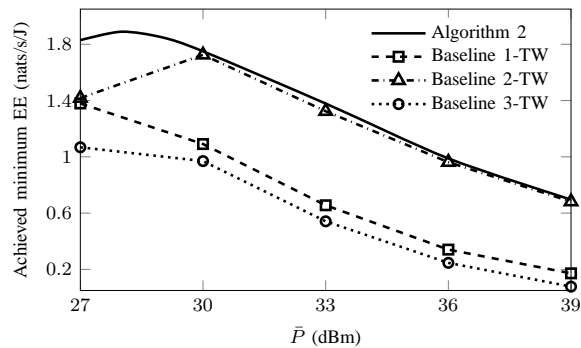}}\vspace{-3mm}

\caption{Performances of Algorithm \ref{Alg.SCA-tw} with $L=12$.}
\vspace{-4mm} \label{figure6}
\end{figure}

In Fig. \ref{figure6}, we evaluate the performances of Algorithm
\ref{Alg.SCA-tw} in terms of convergence and minimum EE. Specifically,
Fig. \ref{fig.6a} plots the convergence behavior of the algorithm
over a random channel realization with two different initial points
also generated randomly. Compared to Algorithm 1, Algorithm \ref{Alg.SCA-tw}
likely requires more iterations to converge. This can be intuitively
explained by the inter-pair interference in two-way relaying systems
which is more difficult to manage than that in one-way relaying systems
due to the bi-directional transmission. Fig. \ref{fig.6b} illustrates
the average achieved minimum EE of Algorithm \ref{Alg.SCA-tw} versus
the maximum transmit power $\bar{P}$. We compare Algorithm \ref{Alg.SCA-tw}
with the three schemes, Baseline 1-TW, Baseline 2-TW and Baseline
3-TW which are set up similarly to Baseline 1-OW, Baseline 2-OW and
Baseline 3-OW in Fig. \ref{Fig.3}. Again, we observe that the proposed
scheme outperforms the others. On the other hand, for Algorithm \ref{Alg.SCA-tw},
we can see that in the region of limited user power, the EE increases
when $\bar{P}$ increases. This is because the effect of the gain
from the additional power resource is stronger than that of the decrease
because of PA efficiency. When $\bar{P}$ is large, an increase of
$\bar{P}$ has insufficient influence, and thus the performance reduces
with $\bar{P}$.

\begin{figure}
\centering{}\subfigure[Average individual EE of user pairs.]{\label{fig. fairness-TW-1}\includegraphics[width=0.47\columnwidth]{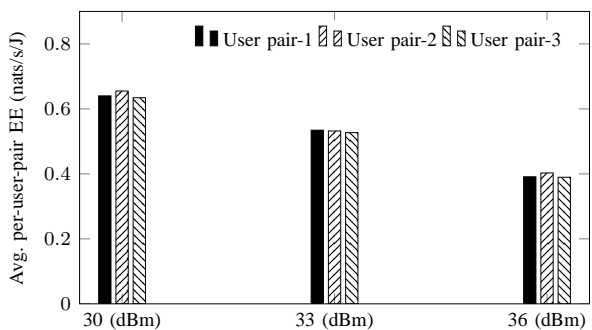}}\quad{}\subfigure[CDF of Jain's fairness index]{\label{fig. fairness-TW-2}\includegraphics[width=0.47\columnwidth]{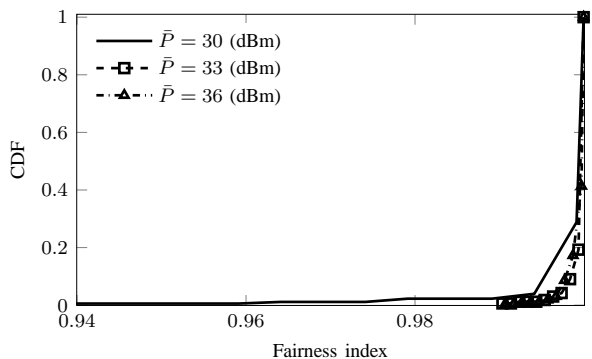}}\vspace{-3mm}\caption{EE fairness among the user pairs achieved by Algorithm 2 with $K=3$
and $L=12$.}
\vspace{-8mm}\label{fig. fairness-TW}
\end{figure}
In Fig. \ref{fig. fairness-TW}, we plot the individual EE performances
of all user pairs (Fig. \ref{fig. fairness-TW-1}) and the CDF of
fairness index (Fig. \ref{fig. fairness-TW-2}) versus different value
of $\bar{P}$. Similar to the observation in Fig. \ref{fig. fairness-OW},
the proposed EE method for two-way relaying is able to maintain the
good EE fairness among all user pairs.\vspace{-5mm}

\subsection{Performance of ZF-Based Designs\label{subsec:Performance-of-ZF}}

\begin{figure}[t]
\centering{}\subfigure[ Average minimum EE of Algorithm 1 and ZF-based scheme in the one-way relaying system.]{\label{fig.7a}\includegraphics[width=0.47\columnwidth]{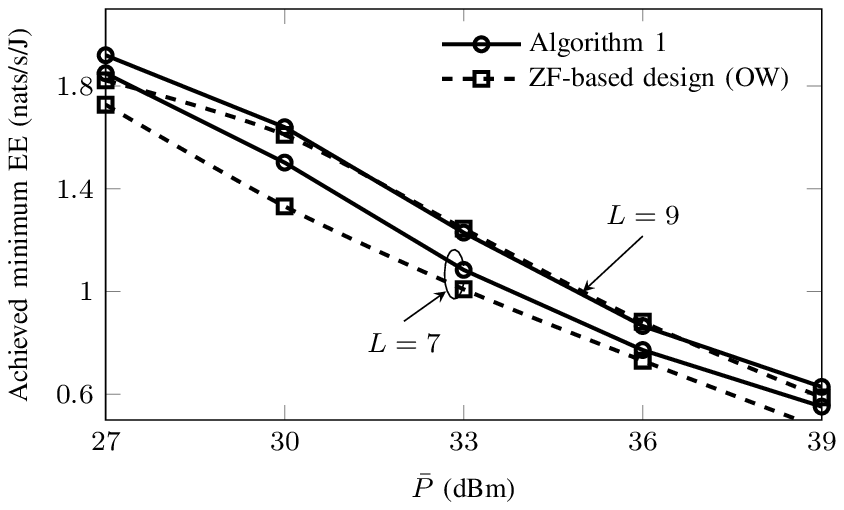}\vspace{-3mm}}\quad{}\subfigure[Average minimum EE of Algorithm 2 and ZF-based scheme in the two-way relaying system.]{\label{fig.7b}\includegraphics[width=0.47\columnwidth]{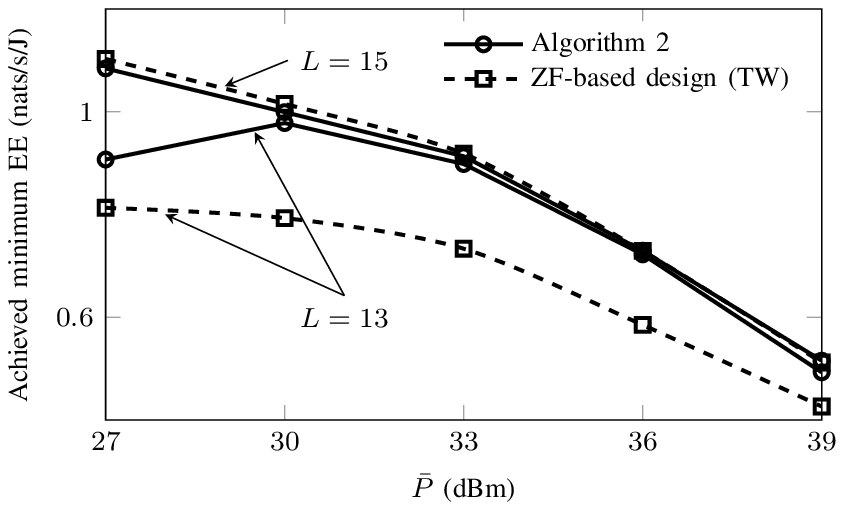}}\vspace{-3mm}\caption{Average minimum EE performances versus $\bar{P}$ of Algorithms 1,
2, and ZF-based schemes.}
\vspace{-4mm}\label{Fig.7}
\end{figure}
In the following set of numerical experiments, we investigate the
performances of ZF-based designs (presented in Section \ref{sec:Computational-Complexity})
in terms of the minimum EE and computational complexity.

Fig. \ref{Fig.7} shows the minimum EE performances of the considered
schemes. In particular, Fig. \ref{fig.7a} plots the performances
of Algorithm 1 and ZF-based design (OW) in the one-way relaying system,
while Fig. \ref{fig.7b} plots the performances of Algorithm 2 and
ZF-based design (TW) in the two-way relaying system. We can observe
that the performances of ZF-based schemes are inferior to Algorithms
1 and 2 when $L$ is small, and comparable, when $L$ is sufficiently
large. The results are because the ZF beamforming needs a certain
number of relays to form the null space.

\begin{figure}
\centering{}\subfigure[ CDF of the number of required iterations for convergence of Algorithms 1 and ZF-based scheme in one-way relaying system.]{\includegraphics[width=0.47\columnwidth]{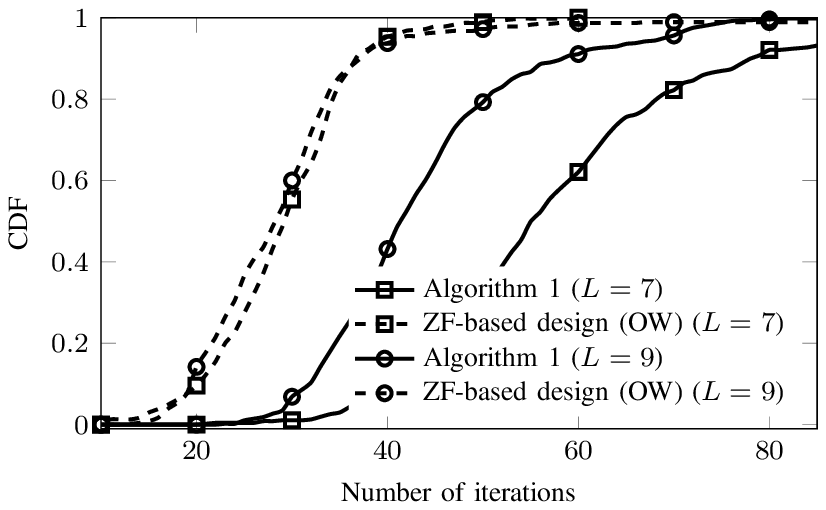}}\quad{}\subfigure[CDF of the number of required iterations for convergence of Algorithms 2 and ZF-based scheme in two-way relaying system.]{\includegraphics[width=0.47\columnwidth]{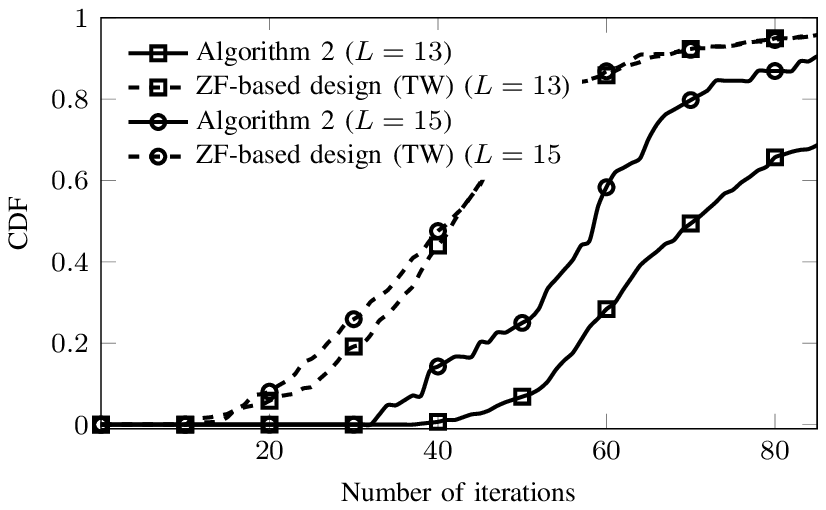}}\vspace{-2mm}\caption{The number of required iterations for convergence of Algorithms 1,
2, and ZF-based schemes.}
\vspace{-8mm}\label{Fig.8}
\end{figure}

To investigate the computational complexity of ZF-based schemes, we
plot in Fig. \ref{Fig.8} the CDFs of the required number of iterations
for convergence of the considered schemes, and provide the corresponding
solver running time in Table 2. It can be observed that the ZF-based
schemes require smaller numbers of iterations to converge compared
to Algorithms 1 and 2. In addition, the solver requires less time
to solve convex subproblems in ZF-based schemes. Consequently, the
total running time of the ZF-based schemes is remarkably smaller than
that of Algorithms 1 and 2. Combining with the results in Fig. \ref{Fig.7},
we can conclude that, when $L$ is large, efficient solutions can
be achieved with low computational cost by using the ZF-based schemes.
\begin{table}
\centering{}\caption{Solver run time (in seconds) for Algorithms 1, 2, and the ZF-based
schemes with $K=3$ and $\bar{P}=33$ $\text{dBm}$.}
\begin{tabular}{c|c|c|c|c||c|c|c|c}
\cline{3-5} \cline{7-9}
\multicolumn{1}{c}{} &  & $L=7$ & $L=8$ & $L=9$ &  & $L=13$ & $L=14$ & $L=15$\tabularnewline
\hline
\multirow{2}{*}{Per-iteration run time} & Algorithm 1 & 0.039 & 0.050 & 0.056 & Algorithm 2 & 0.12 & 0.14 & 0.16\tabularnewline
\cline{2-9}
 & ZF-based (OW) & 0.019 & 0.028 & 0.037 & ZF-based (TW) & 0.021 & 0.026 & 0.030\tabularnewline
\hline
\multirow{2}{*}{Total run time} & Algorithm 1 & 2.24 & 2.53 & 2.40 & Algorithm 2 & 9.97 & 10.16 & 10.36\tabularnewline
\cline{2-9}
 & ZF-based (OW) & 0.56 & 0.82 & 1.08 & ZF-based (TW) & 0.96 & 1.16 & 1.28\tabularnewline
\hline
\end{tabular}\label{Tab2}\vspace{-6mm}
\end{table}
\vspace{-3mm}

\section{Conclusion\label{sec:Conclusion}}

We studied a multipair relay system where the relays harvest energy
from user RF signals. We considered an energy consumption model, which
accounts various realistic aspects such as rate-dependent signal processing
power, dynamic power amplifier efficiency, and nonlinear EH circuits.
We have investigated the problem of max-min EE fairness among user
pairs by jointly designing the transmit data rate, users' transmit
power, relays' processing coefficient, and EH time. For both one-way
and two-way relaying, we have derived iterative procedures based on
the IA optimization framework, where each iteration only deals with
an SOCP. The proposed methods are provably convergent. In addition,
for low-complexity designs, we have proposed an approach based on
a combination of ZF beamforming and IA. The effectiveness of our approaches
has been demonstrated by the numerical results.\vspace{-3mm}

\appendix{}

\subsection{Problem Equivalence\label{sec:Problem-Equivalences}}

We justify the optimal equivalence between (\ref{Prob:epiprob-1})
and EEF-OW as follows. Let us denote $\bm{\psi}^{\ast}$ as the optimal
solution of (\ref{Prob:epiprob-1}) and define $\hat{k}\triangleq\arg\max_{k\in{\cal K}}\{\hat{\eta}_{k}\}$
where $\hat{\eta}_{k}\triangleq\rho_{1k}^{\text{sp}}+\frac{P'_{k}}{r_{1k}^{\ast}}+\frac{\tilde{\tau}^{\ast}}{r_{1k}^{\ast}}\bigl(\frac{\varepsilon_{1k}}{\sqrt{q_{1k}^{\ast}}}+P''_{k}\bigr)$.
We remark that constraints in (\ref{eq:relaypower-1-1}) hold with
equality at the optimum following the epigraph transformation. Thus
it is sufficient to show that: (i) $\hat{\eta}_{\hat{k}}$ is the
optimal solution of (\ref{Prob:epiprob-1}), i.e., $\hat{\eta}_{\hat{k}}=\eta^{\ast}$,
and (ii) (\ref{eq:rate-1}) and (\ref{eq:rate}) with respect to user
pair $\hat{k}$ hold with equality at the optimum. In these regards,
(\ref{Prob:epiprob-1}) and EEF-OW obtain the same optimal values
of $({\bf w}^{\ast},{\bf r}^{\ast},\tau^{\ast},{\bf p}^{\ast}$) as
can be seen by constraints in (\ref{eq:relaypower-1-1}). Thereby,
we achieve $f_{\hat{k}}^{\text{EE,OW}}(\tau^{\ast},{\bf p}^{\ast},{\bf r}^{\ast})=\Bigl(\rho_{1\hat{k}}^{\text{sp}}+\frac{P'_{\hat{k}}}{r_{1\hat{k}}^{\ast}}+\frac{\tilde{\tau}^{\ast}}{r_{1\hat{k}}^{\ast}}\bigl(\frac{\varepsilon_{1\hat{k}}}{\sqrt{q_{1\hat{k}}^{\ast}}}+P''_{\hat{k}}\bigr)\Bigr)^{-1}=\frac{1}{\eta^{\ast}}$
which implies the equivalence between (\ref{Prob:epiprob-1}) and
EEF-OW.

We now show (i). It is immediately seen that (\ref{eq:eesource-equi})
holds at the optimum for $\hat{k}$ and $\hat{\eta}_{\hat{k}}=\eta^{\ast}$.
This is because otherwise $\hat{\eta}_{\hat{k}}<\eta^{\ast}$ which
means that $\eta^{\ast}$ is not the optimum. Next, we prove (ii).
Let us consider problem (\ref{Prob:epiprob-1}) and suppose, to the
contrary, that (\ref{eq:rate-1}) does not hold at the optimum for
$\hat{k}$. Then, we can scale up $r_{1\hat{k}}^{\ast}$ by a positive-scaling
factor $\lambda>1$ such that $\hat{r}_{1\hat{k}}\triangleq\lambda r_{1\hat{k}}^{\ast}=\log(1+v_{1\hat{k}}^{\ast})$.
And, we can easily check that new value $\hat{r}_{1k}$ is still feasible
to (\ref{Prob:epiprob-1}). However, substituting $\hat{r}_{1k}$
to (\ref{Prob:epiprob-1}) results in a strictly smaller objective,
i.e., $\rho_{1\hat{k}}^{\text{sp}}+\frac{P'_{\hat{k}}}{\hat{r}_{1\hat{k}}}+\frac{\tilde{\tau}}{\hat{r}_{1\hat{k}}}\bigl(\frac{\varepsilon_{1\hat{k}}}{\sqrt{q_{1\hat{k}}}}+P''_{\hat{k}}\bigr)=\hat{\eta}_{k}<\eta^{\ast}$.
This contradicts to the fact $\hat{\eta}_{k}=\eta^{\ast}$ at the
optimum. Similarly, we can argue that (\ref{eq:rate}) with respect
to $\hat{k}$ holds with equality at the optimum of EEF-OW. This accomplishes
(ii) and completes the proof.\vspace{-3mm}

\subsection{Convexity of Function $x^{2}/\sqrt{y}$\label{sec:Convexity-of-Function}}

We show that the function is strictly convex over $x>0,y>0$ via the
second-order condition. The Hessian of the function is $\mathbf{A}=\tiny\left[\begin{array}{cc}
2/\sqrt{y} & -x/y^{3/2}\\
-x/y^{3/2} & 3x^{2}/4y^{5/2}
\end{array}\right]$. Then we have
\[
[v_{1}\,v_{2}]\mathbf{A}[v_{1}\,v_{2}]\trans=\frac{2v_{1}^{2}}{\sqrt{y}}-\frac{2xv_{1}v_{2}}{y^{3/2}}+\frac{3x^{2}v_{2}^{2}}{4y^{5/2}}=\frac{2}{\sqrt{y}}(v_{1}-\frac{xv_{2}}{2y})^{2}+\frac{x^{2}v_{1}^{2}}{4y^{5/2}}>0
\]
 for all non-zero vector $[v_{1}\,v_{2}]$, i.e., $\mathbf{A}$ is
positive definite.

It is interesting that the constraint $x^{2}/\sqrt{y}\leq t$ can
be equivalently represented by two SOCs\,as
\[
x^{2}/\sqrt{y}\leq t\Leftrightarrow\begin{cases}
\|[2x,\ t-v]\|_{2}\leq t+v\\
\|[2y,\ v-1]\|_{2}\leq v+1
\end{cases}.
\]
\vspace{-6mm}

\bibliographystyle{IEEEtran}

\end{document}